\documentclass[twocolumn]{article}
\usepackage[misc]{ifsym}
\usepackage[cal=txupr]{mathalpha}

\newcommand{\code}[1]{\texttt{#1}\xspace}

\usepackage{soul, titling, titlesec, enumitem, xspace, placeins, microtype}
\usepackage[margin=1in]{geometry}
\usepackage{flushend}

\usepackage[nodisplayskipstretch]{setspace}
\usepackage[hang,flushmargin,multiple,bottom]{footmisc}
\titlespacing{\section}{0pt}{0.25\baselineskip}{0.1\baselineskip}
\titlespacing{\subsection}{0pt}{0.25\baselineskip}{0.1\baselineskip}
\titlespacing{\subsubsection}{0pt}{0pt}{0pt}

\titleformat*{\section}{\large\bfseries}
\titleformat*{\subsection}{\normalsize\bfseries}

\usepackage{caption, wrapfig, tikz, float, ulem, booktabs, multirow, xcolor, stfloats, colortbl}
\usepackage{circuitikz}
\usetikzlibrary{calc, patterns, patterns.meta}

\let\emph\undefined
\newcommand{\emph}[1]{\textit{#1}}

\definecolor{cornflowerblue}{rgb}{0.39, 0.58, 0.93}
\definecolor{mediumslateblue}{rgb}{0.48, 0.41, 0.93}
\definecolor{saffron}{rgb}{0.96, 0.77, 0.19}

\usepackage{amsthm, cuted}
\newtheorem{theorem}{Theorem}
\newtheorem*{theorem*}{Theorem}
\newtheorem{lemma}[theorem]{Lemma}

\newtheorem*{conjecture*}{Conjecture}

\newtheorem{proposition}[theorem]{Proposition}

\usepackage[many]{tcolorbox}
\newtcolorbox[auto counter]{boxedalgorithm}[2]{%
	label=#2,
	enhanced,
	sharp corners,
	frame hidden,
	center,
	width=\linewidth,
	top=0.5em,
	bottom=0.5em,
	left=0.5em,
	right=0.5em,
	before skip=0.5em,
	after skip=0.5em,
	attach title to upper,
	coltitle=black,
	title={\uline{Algorithm \thetcbcounter\xspace \textit{(#1)}.\hfill}\vspace{2.5pt}},
	breakable
}

\usepackage[many]{tcolorbox}
\newtcolorbox[auto counter]{headlessboxedalgorithm}{%
	enhanced,
	sharp corners,
	frame hidden,
	center,
	width=\linewidth,
	top=0.5em,
	bottom=0.5em,
	left=0.5em,
	right=0.5em,
	before skip=0.5em,
	after skip=0.5em,
	attach title to upper,
	coltitle=black,
	breakable
}

\newlist{enumalgorithm}{enumerate}{2}
\setlist[enumalgorithm,1]{nosep, topsep=1pt, itemsep=2.5pt, label=(\arabic*), leftmargin=*}
\setlist[enumalgorithm,2]{leftmargin=1.5em, topsep=2.5pt, parsep=0pt, itemsep=2.5pt, label=(\alph*), ref=(\arabic{enumalgorithmi}\alph*)}

\newlist{descriptor}{description}{1}
\setlist[descriptor]{labelwidth=0.5in,leftmargin=!,align=left,itemsep=0pt,topsep=0.5em}

\definecolor{amber}{rgb}{1.0, 0.49, 0.0}
\definecolor{amethyst}{rgb}{0.6, 0.4, 0.8}

\usepackage{hyperref}
\usepackage[
	backend=biber,
	style=numeric-comp,
	sorting=noneyear,
	sortcites,
	maxnames=3,
	date=year
]{biblatex}

\DeclareFieldFormat[article,inproceedings,book,online,incollection]{title}{\emph{#1}}

\AtEveryBibitem{\clearfield{pages}}
\AtEveryBibitem{\clearfield{volume}}
\AtEveryBibitem{\clearfield{publisher}}
\AtEveryBibitem{\clearfield{editor}}
\AtEveryBibitem{\clearfield{series}}
\AtEveryBibitem{\clearfield{number}}
\AtEveryBibitem{\clearfield{isbn}}

\AtEveryBibitem{\clearlist{series}}
\AtEveryBibitem{\clearlist{publisher}}
\AtEveryBibitem{\clearlist{editor}}

\DeclareFieldFormat*{url}{}
\DeclareFieldFormat[software]{url}{\mkbibacro{URL}\addcolon\space\url{#1}}

\DeclareFieldFormat*{doi}{}
\DeclareFieldFormat[online,software]{doi}{\mkbibacro{DOI}\addcolon\space\url{#1}}

\DeclareSortingTemplate{noneyear}{
	\sort{\citeorder}
	\sort{\field{year}}
}

\usepackage{amsmath, amsfonts, amssymb, nicefrac, mathtools, mleftright}
\usepackage{quiver}

\allowdisplaybreaks

\providecommand{\F}{\mathbb{F}}
\providecommand{\R}{\mathbb{R}}
\providecommand{\C}{\mathbb{C}}
\providecommand{\T}{\mathbb{T}}
\providecommand{\Z}{\mathbb{Z}}

\let\P\undefined
\DeclareMathOperator{\P}{\mathbf{P}}
\DeclareMathOperator{\K}{\mathcal{K}}

\newcommand{\paren}[1]{\mleft( #1 \mright)}

\newcommand{\abs}[1]{\left\vert#1\right\vert}

\DeclareMathOperator{\rank}{\mathrm{rank}}

\newcommand{\into}{\hookrightarrow}

\let\im\undefined

\DeclareMathOperator{\im}{\mathrm{im}}
\DeclareMathOperator{\Hom}{\mathrm{Hom}}
\newcommand{\betti}{\mathbf{b}}

\let\H\undefined
\DeclareMathOperator{\H}{\mathbf{H}}
\newcommand{\1}{\mathbf{1}}

\DeclareMathOperator{\Energy}{\mathcal{E}}
\DeclareMathOperator{\Occupancy}{\mathcal{N}}

\newcommand{\ATEAMS}{\textsf{ATEAMS}\xspace}
\newcommand{\PHAT}{\textsf{PHAT}\xspace}

\newcommand{\Pangolin}{\textsf{Pangolin}\xspace}

\newcommand{\SparseRREF}{\textsf{SparseRREF}\xspace}
\newcommand{\SpaSM}{\textsf{SpaSM}\xspace}

\newcommand{\D}{\mathcal D}
\newcommand{\sd}{p_{\mathrm{sd}}}

\let\th\undefined
\newcommand{\th}{\textsuperscript{th}\xspace}

\newcommand{\qand}{\quad \text{and} \quad}

\makeatletter
\g@addto@macro \normalsize {%
	\setlength\abovedisplayskip{5pt plus 2pt minus 2pt}%
	\setlength\belowdisplayskip{5pt plus 2pt minus 2pt}%
}
\makeatother

\newcommand{\inverse}[1]{{#1}^{-1}}
\DeclareMathOperator{\Cov}{\mathrm{Cov}}

\DeclareMathOperator{\Var}{\mathrm{Var}}

\newcommand{\curvefit}{\texttt{optimize.curve\_fit}\xspace}

\DeclareMathOperator{\PLGTpartition}{\mathcal Z}
\DeclareMathOperator{\PRCMpartition}{\mathcal Y}

\usepackage{verbatim}

\title{Generalized cluster algorithms for Potts lattice gauge theory}
\def\authorsep{-0.3em}

\author{%
	Anthony E. Pizzimenti $\vcenter{\hbox{\footnotesize\Letter}}$ \\[\authorsep] {\small George Mason University} \\[\authorsep]{\small \texttt{apizzime[at]gmu.edu}} \and%
	Paul Duncan \\[\authorsep] {\small Indiana University} \\[\authorsep] {\small\texttt{pauldunc[at]iu.edu}} \and%
	Benjamin Schweinhart \\[\authorsep] {\small George Mason University} \\[\authorsep] {\small \texttt{bschwei[at]gmu.edu}}%
}

\date{\vspace{0.75em} \textit{\normalsize\today}}

\begin{document}
    \maketitle
    
    \begin{abstract}\noindent   
    	Monte Carlo algorithms, like the \emph{Swendsen--Wang} and \emph{invaded-cluster}, sample the Ising and Potts models asymptotically faster than single-spin Glauber dynamics do. Here, we generalize both algorithms to sample \emph{Potts lattice gauge theory} by way of a $2$-dimensional cellular representation called the \emph{plaquette random-cluster model}. The invaded-cluster algorithm targets Potts lattice gauge theory at criticality by implementing a stopping condition defined in terms of \emph{homological percolation}, the emergence of spanning surfaces on the torus. Simulations for $\Z(2)$ and $\Z(3)$ lattice gauge theories on the cubical $4$-dimensional torus indicate that both generalized algorithms exhibit much faster autocorrelation decay than single-spin dynamics and allow for efficient sampling on $4$-dimensional tori of linear scale at least $40$.
    \end{abstract}

    \section{Introduction}
    	\noindent The \emph{$q$-state Potts lattice gauge theory} (PLGT) on a finite subcomplex $X$ of the cubical complex $\Z^d$ is a random assignment of spins in $\Z(q)$ to the edges of $X$. These assignments are induced by the Hamiltonian \[ \H\paren{f}=-\sum_{x \in X}\1_{\delta^1 f(x) = 1},\] where $x \in X$ is a \emph{plaquette} ($2$-dimensional cell), $\delta^1 f\paren{x}$ is the oriented product of spins on the edges incident to $x$ \cite{wegner,kogut}, and $\Z(q)$ is the multiplicative group of \emph{$q$\th complex roots of unity}. Recent work in~\cite{bernoulli-cell-complexes,duncan2023sharp,shklarov} generalizes the Fortuin--Kastelyn \emph{random-cluster model} \cite{fk-random-cluster} (a graphical representation of the Potts model) to construct a $2$-dimensional cellular representation of Potts lattice gauge theory called the \emph{plaquette random-cluster model} (PRCM). We use this cellular representation to define non-local Monte Carlo algorithms for Potts lattice gauge theory. Specifically, we generalize the \emph{Swendsen--Wang}~\cite{sw} and \emph{invaded-cluster} algorithms~\cite{ic,ic-eq}, procedures that revolutionized computational simulations of the Potts model. Both algorithms alternate between sampling $2$-subcomplexes and sampling spins: the former does so according to the PRCM and PLGT;\footnote{An algorithm of this form was previously described for the special case $q=2$ in~\cite{brower1990plaquette,ben1990critical}.} the latter builds a $2$-complex cell-by-cell until a stopping condition based on \emph{homological percolation}  --- the emergence of ``giant surfaces'' \---- is met, then re-samples spins and begins anew.

        \begin{figure}
	\centering
	\begin{circuitikz}
		\def\width{0.5}
		\def\opacity{0.5}

		\tikzset{
			closed/.style={ultra thick, dashed},
			open/.style={ultra thick},
			square/.style={shape=rectangle, minimum size=\width in, fill opacity=\opacity, draw=none},
			supertinylabel/.style={font=\tiny, inner sep=2pt},
			label/.style={font=\small, inner sep=3pt},
			biglabel/.style={font=\footnotesize, inner sep=6pt},
			tinylabel/.style={font=\footnotesize, inner sep=5pt},
			circ/.style={draw,circle,fill=black,inner sep=1pt},
			middy/.style={solid,currarrow,pos=0.5,sloped,scale=0.75}
		}
		
		\node[square, fill=gray] (left) at (0,0) {};
		\node[square, anchor=west] (right) at (left.east) {};
		\node[square, anchor=north, fill=gray] (bottomleft) at (right.south) {};
		\node[square, anchor=west, fill=gray] (bottomright) at (bottomleft.east) {};
		\node[square, anchor=east] (moreleft) at (left.west) {};
		\node[square, anchor=north,fill=gray] (waymoreleft) at (moreleft.south) {};
		
		\foreach \square in {left,right,bottomleft,waymoreleft} {
			\draw[open] (\square.north west) -- (\square.north east) node[middy] {};
			\draw[closed] (\square.north west) -- (\square.south west) node[middy,xscale=-1] {};
			\draw[closed] (\square.south west) -- (\square.south east) node[middy] {};
			\draw[open] (\square.south east) [open]-- (\square.north east) node[middy] {};
		}
		
		\foreach \square in {left,right,bottomleft,bottomright,moreleft,waymoreleft} {
			\foreach \anchor in {north west,north east,south west,south east} {
				\node[circ] at (\square.\anchor) {};
			}
		}
		
		\draw[closed] (bottomright.north west) -- (bottomright.north east) node[middy] {};
		\draw[open] (bottomright.north west) -- (bottomright.south west) node[middy,xscale=-1] {};
		\draw[open] (bottomright.south west) -- (bottomright.south east) node[middy] {};
		\draw[closed] (bottomright.south east) -- (bottomright.north east) node[middy] {};
		
		\draw[closed] (moreleft.north west) -- (moreleft.north east) node[middy] {};
		\draw[open] (moreleft.north west) -- (moreleft.south west) node[middy,xscale=-1] {};
		\draw[open] (moreleft.south west) -- (moreleft.south east) node[middy] {};
		\draw[closed] (moreleft.south east) -- (moreleft.north east) node[middy] {};

		\node[label, anchor=north west] at (waymoreleft.north west) {\mathstrut$c$};
		\node[label, anchor=north east] at (waymoreleft.north east) {$d$};
		\node[label, anchor=south west] at (waymoreleft.south west) {$a$};
		\node[label, anchor=south east] at (waymoreleft.south east) {$b$};
	\end{circuitikz}
	\caption{A cubical complex with fourteen $0$-cells (vertices), nineteen $1$-cells (edges), and four $2$-cells (filled squares). Edge coefficients are $0$ (solid) or $1$ (dashed).}
	\label{figure:edge-labeled-cubical-complex}
\end{figure}
        
        In Section~\ref{section:introduction}, we define Potts lattice gauge theory, the plaquette random-cluster model, and the coupling between them. Section~\ref{section:algorithms} covers the Swendsen--Wang and invaded-cluster algorithms, including a proof that the former has the correct stationary distribution. We then discuss consequences of the duality properties of the PRCM in Section~\ref{section:duality}; the Loop--Cluster coupling~\cite{zhang2020loop} and its generalization~\cite{hansen2025general} are recovered by considering the behavior of the coupling between Potts (hyper)lattice gauge theory and the plaquette random-cluster model under the latter's duality transformation. In Section~\ref{section:experiments}, we conclude by describing practical implementations of the Swendsen--Wang and invaded-cluster algorithms and analyzing autocorrelation decay in computational simulations.
        
        

    \section{Background and definitions}\label{section:introduction}

            \noindent A \emph{finite cubical complex} $X$ is a combinatorial object made up of vertices, edges, and \emph{plaquettes} (squares). Typically, $X$ is a finite subset of $\Z^d$, the tiling of $\R^d$ by unit $d$-cubes with corners at integer coordinates. Elements of $\Z^d$  are \emph{$i$-cells}, where $i$ indicates dimension: $0$-cells are vertices of the integer lattice, $1$-cells are edges in the nearest-neighbors graph on the vertices, and $2$-cells the (square) minimal cycles formed by the edges. In this context, ``$2$-cells,'' ``plaquettes,'' and ``$2$-plaquettes'' are the same. We are mostly interested in cells of dimension less than $3$ (though $3$-cells enter into the definition of the plaquette invaded-cluster algorithm). To impose periodic boundary conditions on finite complexes, we ``glue'' opposite cells of $[0,N]^d \subset \Z^d$ together to form the \emph{discrete torus} $\T^d_N$. See Figure \ref{figure:edge-labeled-cubical-complex} for an illustration.
			
		\subsection{Potts lattice gauge theory}\label{section:introduction:subsection:PLGT}
            \noindent The \emph{$q$-state Potts lattice gauge theory} (PLGT) on a cubical complex $X$ assigns spins in $\Z(q)$ to the edges of $X$. 
            A spin assignment to the edges of $X$ is called a \emph{$1$-cochain} or a \emph{discrete $1$-form}; we denote the group of these assignments by $C^1(X;\Z(q))$. Each $f \in C^1(X;\Z(q))$ is a function that assigns a spin to each edge of $X$ so that the spin is inverted whenever its corresponding edge's orientation is reversed. When $q$ is prime, $C^1(X;\Z(q))$ is a $\Z(q)$-vector space. Potts lattice gauge theory has been studied as a relatively simple discrete example of a model with a gauge symmetry: when $q=2$ or $q=3$, it coincides with $\Z\paren{q}$ \emph{Euclidean lattice gauge theory}. In addition, \emph{Ising lattice gauge theory} --- the case $q=2$ --- has met renewed interest because of its relationship with the \emph{toric code}~\cite{agrawal2024geometric} from quantum information theory, spurring on development of fast-converging algorithms for sampling these models.
			
			The \emph{coboundary} (or \emph{discrete exterior derivative}) of a $1$-cochain is a $2$-cochain --- a map assigning $\Z(q)$ spins to plaquettes --- and belongs to the group $C^2(X;\Z(q))$ of $2$-cochains on $X$. For $f \in C^1(X;\Z(q))$ and a plaquette $x \in X$, the coboundary $\delta^1 f$ assigns to $x$ the oriented product of the spins on its bounding edges. The plaquette $x = \{abdc\}$ in Figure \ref{figure:edge-labeled-cubical-complex}, for example, is oriented so that its boundary is \[ \partial_2(x) = \{ab\} \inverse{\{ac\}} \{bd\} \inverse{\{cd\}}.\] Extending multiplicatively, we define $(\delta^1 f)(x)$ by
			\begin{align*}
				(\delta^1 f) (x) &\coloneqq f(\{ab\}) \inverse{f(\{ac\})} f(\{bd\}) \inverse{f(\{cd\})} \\
				&= f(\partial_2(x)).
			\end{align*}
            We can also take the coboundary of a vertex spin assignment $g\in C^0(X;\Z(q)).$ If $x$ is an oriented edge from $v$ to $w$, then $(\delta^0 g)\paren{x}=g\paren{w} \inverse{g\paren{v}}.$ As $\delta^1\circ\delta^0=1$, the Hamiltonian is unchanged by the addition of the coboundary of a $0$-cochain: this is the action of gauge transformations on $C^1\paren{X;\Z(q)}$. See Appendix \ref{appendix:definitions} for complete definitions.
            
			The $q$-state Potts lattice gauge theory $\nu$ on $X$ with inverse temperature $\beta$ is then the Gibbs distribution on $C^1\paren{X;\Z(q)}$ induced by the Hamiltonian $\H$ that counts the number of non-frustrated plaquettes $x \in X$ (i.e. plaquettes $x$ such that $\delta^1 f(x) = 1$)~{\cite{wegner,kogut}} given by \[\allowdisplaybreaks \H\paren{f}=-\sum_{x \in X}\1_{\delta^1 f(x) = 1}. \] In all, PLGT has law \[\nu(f) = \frac{1}{\PLGTpartition(q,\beta)} e^{-\beta \H(f)}, \] assigning the probability $\nu\paren{f}$ to the edge spin assignment $f$. We can extend this model to define a random $i$-cochain in $C^i\paren{X;\Z(q)}$, where the case $i=0$ is the classical $q$-state Potts model \cite{wegner,bernoulli-cell-complexes}.

		\subsection{Plaquette random-cluster model}
			\noindent To derive the \emph{plaquette random-cluster model}, we use the same strategy as for the random-cluster model: expand and manipulate the partition function of its corresponding Potts model. For context, we first divine the classical random-cluster model. If $G$ is a finite graph with vertices $V_G$ and edges $E_G$, $C^0(G;\Z(q))$ is the set of spin assignments $f: V_G \to \Z(q)$. Given a subgraph $P \subset G$, an assignment $f$ is \emph{compatible} with $P$ whenever $f(u) = f(w)$ for all edges $(v,w)$ in $P$; call the set of all compatible spin assignments $Z^0(P;\Z(q))$. With $\beta$ an inverse temperature parameter, $p = 1-e^{-\beta}$, and $F(f,x)$ the event that $f(v) = f(w)$ for the edge $x = (v,w)$ in $P$,
				\begin{align*}
					e^{\beta \1_{F(f,x)}} &= e^\beta \left( (1-e^{-\beta}) \1_{F(f,x)} + e^{-\beta} \right) \\
					&= e^\beta\left( p \1_{F(f,x)} + (1-p) \right)
				\end{align*}
				with $f$ fixed. Noticing that \[ \sum_f e^{-\beta \H(f)} = \sum_f \prod_{x} e^{\beta \1_{F(f,x)}} \] over edges $x$ in $G$, the partition function $\PLGTpartition \coloneqq \PLGTpartition(q,\beta)$ for the Potts model on $G$ is given by 
				\begin{align*}
					\PLGTpartition &= \sum_f \prod_{x} e^\beta\left( p \1_{F(f,x)} + (1-p) \right) \\
					&= e^{\beta \abs{E_G}} \left[\sum_{P \subset G} p^{\abs{E_P}}(1-p)^{\abs{E_G}-\abs{E_P}} \abs{Z^0(P; \Z(q))} \right]{.}
				\end{align*} Spin assignments are compatible with $P$ exactly when they are constant on the connected components of $P$, so $\abs{Z^0(P; \Z(q))}$ equals $q^{\betti_0(P)},$ where $\betti_0(P)$ denotes the number of components\footnote{This equals the size of the zero-dimensional homology group $H_0\paren{P,\Z(q)}$.}. The \emph{random-cluster model} assigns $P$ probability proportional to \[ p^{\abs{E_P}}(1-p)^{\abs{E_G}-\abs{E_P}} q^{\betti_0(P)}, \] with partition function $\PRCMpartition(p,q) \coloneqq e^{-\beta \abs{E_G}} \PLGTpartition(q,\beta)$~\cite{rcm}.
            
			In the \emph{plaquette random-cluster model} we swap the Potts model for PLGT, increasing dimension as needed. Given a finite cubical complex $X$ with vertices $V_X$, edges $E_X$, and plaquettes $U_X$, a \emph{percolation subcomplex} $P$ is a subcomplex of $X$ where $V_P = V_X$, $E_P = E_X$, and $U_P \subset U_X$. A spin assignment $f \in C^1(X;\Z(q))$ is \emph{compatible} with $P$ when, for all plaquettes $x \in U_P$, the oriented product of spins on the edges of $x$ is $1$ --- i.e. when $\delta^1 f (x) = 1$. Like before, the set of compatible assignments is written as $Z^1(P;\Z(q))$, and $F(f,x)$ is the event that $\delta^1 f(x) = 1$. Using the same trick, the partition function $\PLGTpartition \coloneqq \PLGTpartition(q,\beta)$ for Potts lattice gauge theory is
			\begin{align*}
				\PLGTpartition &= \sum_f \prod_{x} e^\beta\left( p \1_{F(f,x)} + (1-p) \right) \\
				&= e^{\beta \abs{U_X}} \left[\sum_{P \subset X} p^{\abs{U_P}}(1-p)^{\abs{U_X}-\abs{U_P}} \abs{Z^1(P; \Z(q))} \right].
			\end{align*}
			
			\noindent Here, there is a key difference from the $0$-dimensional case: we can no longer in general express $\abs{Z^1(P; \Z(q))}$ as $q$ to a power independent of $q$ and, when $q$ is not prime, it may not be a power of $q$ at all. Before explaining why, we account for the action of the gauge transformation on $C^1(X;\Z(q))$. Any vertex spin assignment $g \in C^0(X;\Z(q))$ gives rise to an edge spin assignment $\delta^0 g\in C^1\paren{X;\Z(q)}.$ As consequences of the identity $\delta^1 \circ \delta^0=1$, we have that $g\in Z^1\paren{P;\Z(q)}$ for any $P$, and the Hamiltonian $\H$ of PLGT remains invariant under the gauge transformation $f\mapsto f \cdot \delta^0 g$. The group of \emph{$1$-coboundaries} $B^1(P;\Z(q))$ is the collection of gauge transformations $\delta^0 g$; importantly, these do not depend on $P$. The quotient group \[ H^1(P;\Z(q)) \coloneqq Z^1(P;\Z(q))/B^1(P;\Z(q)) \] of cocycles modulo coboundaries is the \emph{cohomology} of $P$. (See Appendix \ref{appendix:cohomological-rank} for an example computation.) By Lagrange's theorem on quotient groups, \[ \abs{Z^1(P;\Z(q))} = \abs{H^1(P;\Z(q))} \abs{B^1(X;\Z(q))}. \] Abbreviating $B^1(X;\Z(q))$ and $H^1(P;\Z(q))$ to $B^1_q$ and $H^1_q$ respectively, we can rewrite $\PLGTpartition(q,\beta)$ as \[ e^{\beta \abs{U_X}} \abs{B^1_q} \left[\sum_{P \subset X} p^{\abs{U_P}}(1-p)^{\abs{U_X}-\abs{U_P}} \abs{H^1_q} \right].\] The \emph{plaquette random-cluster model} (PRCM) $\mu$ with parameters $p=1-e^{-\beta}$ and $q$ assigns the percolation subcomplex $P \subset X$ probability \[ \mu\paren{P}=\frac{1}{\PRCMpartition \paren{p,q}}p^{\abs{U_P}}\paren{1-p}^{\abs{U_X}-\abs{U_P}}\abs{H^1\paren{P;\Z(q)}}. \] Akin to the $0$-dimensional case, \[\PRCMpartition(p,q) = \frac{e^{-\beta |U_X|}}{\abs{B^1(X;\Z(q))}}\PLGTpartition(q,\beta). \]
			
			Hiraoka and Shirai introduced the prime-$q$ case of the PRCM in the context of a mean-field analogue of this model, but did not connect their work to the literature on Potts lattice gauge theory \cite{bernoulli-cell-complexes}. Duncan and Schweinhart further studied the PRCM on $\Z^d$ in \cite{prcm-plgt}: they proved a formula equating Wilson loop expectations in PLGT with the probability of a corresponding topological event in the PRCM, and established the the existence of phase transitions defined in terms of homological percolation (see Theorem~\ref{theorem:ds-8} below). The extension to general $q$ was independently discovered by Duncan and Schweinhart \cite{duncan2023sharp,prcm-24} and Shklarov \cite{shklarov}; the former authors established a rigorous relationship between the deconfinement transition for PLGT on $\Z^3$ and the percolation transition for the random-cluster model. Earlier attempts at this construction~\cite{ginsparg1980large,maritan1982gauge} fell short~\cite{aizenman1984topological} because they implicitly assumed that $|H^1(P;\Z(q))|$ always equals $q^{\rank H^1(P;\Z(q))}$, but --- even when $q$ is prime --- $\left| H^1\left(P;\Z(q)\right)\right|$ depends on $q$. Though $\abs{H^1(P;\Z(q))}$ is a power of $q$ when $P$ is a subcomplex of $\Z^3$, this fails in higher dimensions and also for the $3$-torus. If $P\subset\Z^4$ is an embedding of the real projective plane, for example, $\left| H^1\left(P;\Z_2\right)\right|=2$ but $\left| H^1\left(P;\Z(q)\right)\right|=0$ for odd $q$. There are more complicated examples for non-prime $q$ where $\abs{H^1\left(P;\Z(q)\right)}$  is not a power of $q.$ By contrast, the size of the $0$-dimensional cohomology $\abs{H^0\paren{P;\Z(q)}}$ always equals $q^{\betti_0(P)}$. See \cite{aizenman1984topological} for more discussion of this phenomenon.

		\subsection{Coupling}
			\noindent Following Edwards and Sokal \cite{edwards-sokal}, we couple Potts lattice gauge theory $\nu$ (with parameters $\beta$ and $q$) to the PRCM $\mu$ (with parameters $p=1-e^{-\beta}$ and $q$) as the joint distribution $\K$ assigning probability \[ \K(f,P) \propto \prod_{\mathclap{x \in X}} \left[ (1-p)\left(\1_{A(x)}\right) + p\left(\1_{B(f,x)}\right)\right] \] to each pair of edge spin assignment $f\in C^1\paren{X;\Z(q)}$ and percolation subcomplex $P\subset X$. The product is taken over the plaquettes $x$ of $X$: $A(x)$ is the event that $x \not \in P$, and $B(f,x)$ the event that $x \in P$ \emph{and} $\delta^1 f(x) = 1$. The coupling is illustrated (in an algorithmic context) in Figure \ref{figure:sw-illustrated}. Based on work by Hiraoka and Shirai in \cite{bernoulli-cell-complexes}, Duncan and Schweinhart prove the following as Proposition 12 and Corollary 13 in \cite{duncan2023sharp}:
			
			\vspace{0.5em}
			\begin{proposition}[Marginals of $\K$, Prop. 12 in~\cite{duncan2023sharp}]\label{proposition:marginals}
				The marginal distribution $\K(f, -)$ on $f$ is Potts lattice gauge theory, and the marginal distribution $\K(-, P)$ on $P$ is the plaquette random-cluster measure.
			\end{proposition}
			\vspace{0.5em}
			
			\begin{proposition}[Conditionals of $\K$, Cor. 13 in \cite{duncan2023sharp}]\label{proposition:conditionals}
				The conditional distribution $\K(- \mid f)$ on percolation subcomplexes $P$ given a cochain $f$ is independent percolation with probability $p$ over non-frustrated plaquettes in $X$. The conditional distribution $\K(- \mid P)$ on cochains $f$ given a percolation subcomplex $P$ is the uniform measure on $Z^1 \paren{P;\Z(q)}=\ker\delta^1$, the group of cocycles of $P$.
			\end{proposition}
			\vspace{0.5em}
			

    \section{Cluster algorithms for PLGT}
    	\label{section:algorithms}

\begin{figure*}[hb]
	\centering
	\vspace{-0.5em}
	\begin{tikzpicture}
		\def\opacity{0.6}						
		\def\L{1.25}								
		\def\shift{4.5}							
		\def\linewidth{1.75pt}						
		\def\exploded{0.4}						
		\def\sep{0.2in}							
		\def\direction{1}						
		\def\labels{1}
		
		\ifthenelse{\direction < 1}{
			\def\xshift{0}
			\def\yshift{-\shift}
			
			\def\xlabelsep{\sep}
			\def\ylabelsep{0}
			\def\labelanchor{east}
			\def\rotation{270}
			
			\def\labelfrom{south}
			\def\labelto{north}
		}{
			\def\yshift{0}
			\def\xshift{\shift}
			
			\def\xlabelsep{0}
			\def\ylabelsep{-\sep}
			\def\labelanchor{south}
			\def\rotation{0}
			
			\def\labelfrom{east}
			\def\labelto{west}
		}
		
		\tikzset{
			open/.style={fill=gray, fill opacity=\opacity, draw=none},
			closed/.style={fill=none, draw=none},
			none/.style={fill=none, draw=none},
			o/.style={line width=\linewidth,black,line cap=round},
			c/.style={line width=\linewidth,black, dash pattern={on 3pt off 2pt}},
			n/.style={draw=none},
			steplabel/.style={font={\tiny\bfseries\color{white}}, draw=gray, fill=gray, circle, line width=2pt, inner sep=0.5pt},
			steplabelline/.style={gray,line width=1pt,-{Latex[length=5pt]}},
			steplabellinelabel/.style={midway,fill=white,font={\footnotesize},rotate=\rotation}
		}
		
		\coordinate (0) at (0,0,0);
		\coordinate (1) at (\L,0,0);
		\coordinate (2) at (0,\L,0);
		\coordinate (3) at (\L,\L,0);
		\coordinate (4) at (0,0,-\L);
		\coordinate (5) at (\L,0,-\L);
		\coordinate (6) at (0,\L,-\L);
		\coordinate (7) at (\L,\L,-\L);
		
		\begin{scope}[shift={(-\xshift,-\yshift)},ec/.style={shift={(-\xshift,-\yshift)}}, local bounding box=step-1]
			\def\Plaquettes{{%
				{0/4/6/2/open/o/c/n/c},%
				{4/5/7/6/open/c/o/n/n},%
				{2/3/7/6/open/o/o/o/o},%
				{0/1/5/4/closed/c/c/n/n},%
				{0/1/3/2/closed/n/o/n/n}%
			}}
			
			\foreach \F in \Plaquettes {
				\foreach \p/\q/\r/\s/\style/\a/\b/\c/\d in \F {
					\draw[none] ([ec]\p) -- ([ec]\q) -- ([ec]\r) -- ([ec]\s) -- cycle;
					\draw[\a] ([ec]\p) -- ([ec]\q);
					\draw[\b] ([ec]\q) -- ([ec]\r);
					\draw[\c] ([ec]\r) -- ([ec]\s);
					\draw[\d] ([ec]\s) -- ([ec]\p);
				}
			}
		\end{scope}

		\begin{scope}[shift={(0,0)},ec/.style={shift={(0,0)}}, local bounding box=step-2]
			\def\Plaquettes{{%
				{4/5/7/6/closed/c/o/o/c/0/0/-\exploded},
				{0/4/6/2/open/o/c/o/c/-\exploded/0/0},
				{0/1/5/4/closed/c/c/c/o/-0.25*\exploded/-\exploded/0},
				{1/5/7/3/closed/c/o/o/o/\exploded/0/0},
				{2/3/7/6/open/o/o/o/o/-0.25*\exploded/\exploded/0},
				{0/1/3/2/open/c/o/o/c/0/0/\exploded}
			}}
			
			\foreach \F in \Plaquettes {
				\foreach \p/\q/\r/\s/\style/\a/\b/\c/\d/\x/\y/\z in \F {
					\draw[\style] ([ec]$(\p)+(\x,\y,\z)$) -- ([ec]$(\q)+(\x,\y,\z)$) -- ([ec]$(\r)+(\x,\y,\z)$) -- ([ec]$(\s)+(\x,\y,\z)$) -- cycle;
					\draw[\a] ([ec]$(\p)+(\x,\y,\z)$) -- ([ec]$(\q)+(\x,\y,\z)$);
					\draw[\b] ([ec]$(\q)+(\x,\y,\z)$) -- ([ec]$(\r)+(\x,\y,\z)$);
					\draw[\c] ([ec]$(\r)+(\x,\y,\z)$) -- ([ec]$(\s)+(\x,\y,\z)$);
					\draw[\d] ([ec]$(\s)+(\x,\y,\z)$) -- ([ec]$(\p)+(\x,\y,\z)$);
				}
			}
		\end{scope}
		
		\begin{scope}[shift={(\xshift,\yshift)}, ec/.style={shift={(\xshift,\yshift)}}, local bounding box=step-3]
			\def\Plaquettes{{%
				{4/5/7/6/closed/o/o/c/c/0/0/-\exploded},
				{0/4/6/2/open/c/c/c/c/-\exploded/0/0},
				{0/1/5/4/closed/o/o/o/c/-0.25*\exploded/-\exploded/0},
				{1/5/7/3/closed/o/o/o/c/\exploded/0/0},
				{2/3/7/6/open/o/o/c/c/-0.25*\exploded/\exploded/0},
				{0/1/3/2/open/o/c/o/c/0/0/\exploded}
			}}
			
			\foreach \F in \Plaquettes {
				\foreach \p/\q/\r/\s/\style/\a/\b/\c/\d/\x/\y/\z in \F {
					\draw[\style] ([ec]$(\p)+(\x,\y,\z)$) -- ([ec]$(\q)+(\x,\y,\z)$) -- ([ec]$(\r)+(\x,\y,\z)$) -- ([ec]$(\s)+(\x,\y,\z)$) -- cycle;
					\draw[\a] ([ec]$(\p)+(\x,\y,\z)$) -- ([ec]$(\q)+(\x,\y,\z)$);
					\draw[\b] ([ec]$(\q)+(\x,\y,\z)$) -- ([ec]$(\r)+(\x,\y,\z)$);
					\draw[\c] ([ec]$(\r)+(\x,\y,\z)$) -- ([ec]$(\s)+(\x,\y,\z)$);
					\draw[\d] ([ec]$(\s)+(\x,\y,\z)$) -- ([ec]$(\p)+(\x,\y,\z)$);
				}
			}
		\end{scope}

		\begin{scope}[shift={(2*\xshift,2*\yshift)},ec/.style={shift={(2*\xshift,2*\yshift)}}, local bounding box=step-4]
			\def\Plaquettes{{%
				{0/4/6/2/open/c/n/n/c},%
				{4/5/7/6/open/o/o/n/c},%
				{2/3/7/6/open/o/o/c/c},%
				{0/1/5/4/closed/o/o/n/n},%
				{0/1/3/2/closed/n/c/n/n}%
			}}
			
			\foreach \F in \Plaquettes {
				\foreach \p/\q/\r/\s/\style/\a/\b/\c/\d in \F {
					\draw[none] ([ec]\p) -- ([ec]\q) -- ([ec]\r) -- ([ec]\s) -- cycle;
					\draw[\a] ([ec]\p) -- ([ec]\q);
					\draw[\b] ([ec]\q) -- ([ec]\r);
					\draw[\c] ([ec]\r) -- ([ec]\s);
					\draw[\d] ([ec]\s) -- ([ec]\p);
				}
			}
			
		\end{scope}
		
		\ifthenelse{\labels > 0}{
			\node[steplabel] (step-2-label) at ($(step-2.\labelanchor)+(\xlabelsep,\ylabelsep)$) {2};
			\node[steplabel] (step-1-label) at ($(step-2-label)+(-\xshift,-\yshift)$) {1};
			\node[steplabel] (step-3-label) at ($(step-2-label)+(\xshift,\yshift)$) {3};
			\node[steplabel] (step-4-label) at ($(step-3-label)+(\xshift,\yshift)$) {4};
			
			\draw[steplabelline] (step-1-label.\labelfrom) -- (step-2-label.\labelto) node[steplabellinelabel] {$P_{t+1} \sim \K(- \mid f_t)$};
			
			\draw[steplabelline] (step-2-label.\labelfrom) -- (step-3-label.\labelto) node[steplabellinelabel] {$f_{t+1} \sim \K(- \mid P_{t+1})$};
			
			\draw[steplabelline] (step-3-label.\labelfrom) -- (step-4-label.\labelto) node[steplabellinelabel] {return $f_{t+1}$};
		}{}
	\end{tikzpicture}
	\vspace{-3pt}
	\caption{The plaquette Swendsen--Wang algorithm at work for $q=2$; edge coefficients are $1$ (solid) or $-1$ (dashed). Given some initial cochain $f_t$, the algorithm first samples $P_{t+1}$ conditioned on $f_t$, then $f_{t+1}$ conditioned on $P_{t+1}$. The figure also illustrates the coupled measure $\K$ on the space of cochains and subcomplexes.}
	\label{figure:sw-illustrated}
\end{figure*}

			\noindent Using the coupling, we can generalize the \emph{Swendsen--Wang algorithm} \cite{sw}. Given parameters $p=1-e^{-\beta}$ and $q\in\Z_{\geq 2}$, the following procedure takes as input the spin assignment $f_{t}\in C^1\paren{X;\Z(q)}$ at time $t$, and outputs an updated cochain $f_{t+1}$ after passing through a percolation subcomplex $P_{t+1}$. As we prove below, the sample distributions for $f_t$ and $P_t$ converge to PLGT and the PRCM, respectively.  
			
\begin{boxedalgorithm}{Plaquette Swendsen--Wang (PSW)}{algorithm:psw}


	\begin{enumalgorithm}
		\item \label{algorithm:psw:step:omega} For each plaquette $x \in X$, sample $\alpha$ uniformly over $[0,1]$ and set \vspace{-0.5em} 
			\[ \omega_{t+1}(x) \coloneqq
				\begin{cases}
					0 & \delta^1 f_t(x) \neq 1 \\
					\begin{cases}
						1 & \alpha < p \\
						0 & \alpha \geq p
					\end{cases} & \delta^1 f_t(x) = 1
				\end{cases}
			\]\vspace{-1em}
		\item\label{algorithm:psw:step:subcomplex} Let $P_{t+1}$ comprise all vertices and edges of $X$, and all plaquettes $x$ where $\omega_{t+1}(x)=1$.
		\item \label{algorithm:psw:step:sample} Uniformly sample $f_{t+1}$ from $Z^1\paren{P_{t+1};\Z(q)}$, and set $t \coloneqq t+1$.
		\item Return to \ref{algorithm:psw:step:omega} unless a suitable convergence threshold is met.
	\end{enumalgorithm}
\end{boxedalgorithm}
			
			\begin{lemma}\label{lemma:irreducible-aperiodic}
			    The Markov chain $\mathcal M$ simulated by the plaquette Swendsen--Wang algorithm (Algorithm \ref{algorithm:psw}) is irreducible and aperiodic, so it has a unique stationary distribution.
			\end{lemma}
			\vspace{0.5em}
			
			\noindent Lemma \ref{lemma:irreducible-aperiodic} immediately implies the following theorem.
			
			\vspace{0.5em}
			\begin{theorem}\label{theorem:invariant-measure}
			    The $q$-state Potts lattice gauge theory $\nu$ is a reversible distribution for $\mathcal M$, and so is the unique stationary distribution for $\mathcal M$.
			\end{theorem}
			\vspace{-1em}
			\begin{proof}\allowdisplaybreaks
			    Let $f$ and $f^*$ be cochains. Algorithm \ref{algorithm:psw} tells us that the probability $\P(f^* \mid f)$ of sampling $f^*$ given $f$ is the product of the conditional distributions in Proposition \ref{proposition:conditionals}. With Proposition \ref{proposition:marginals}'s assertion that $\K(f, -) = \nu(f)$, we can say that
			    \begin{align*}
			        \nu(f) \P(f^* \mid f) &= \K(f,-) \sum_P \K(f^* \mid P) \K(P \mid f) \\
			        &= \sum_P \K(f, -) \frac{\K(f^*, P)}{\K(-, P)} \frac{\K(f, P)}{\K(f, -)} \\
			        &= \sum_P \K(f^*, -) \frac{\K(f, P)}{\K(-, P)} \frac{\K(f^*, P)}{\K(f^*, -)} \\
			        &= \K(f^*,-) \sum_P \K(f \mid P) \K(P \mid f^*) \\
			        &= \nu(f^*)\P(f \mid f^*),
			    \end{align*}
                so $\nu$ is a reversible --- and thus stationary --- distribution for $\mathcal M$. By Lemma \ref{lemma:irreducible-aperiodic}, Potts lattice gauge theory is the unique stationary distribution for $\mathcal M$.
			\end{proof}
			
\begin{figure}[t]
    \centering
    \begin{tikzpicture}

        \node (surface) at (0,0) {\includegraphics[trim={0 0.9in 0 0.9in},clip,width=0.95\linewidth]{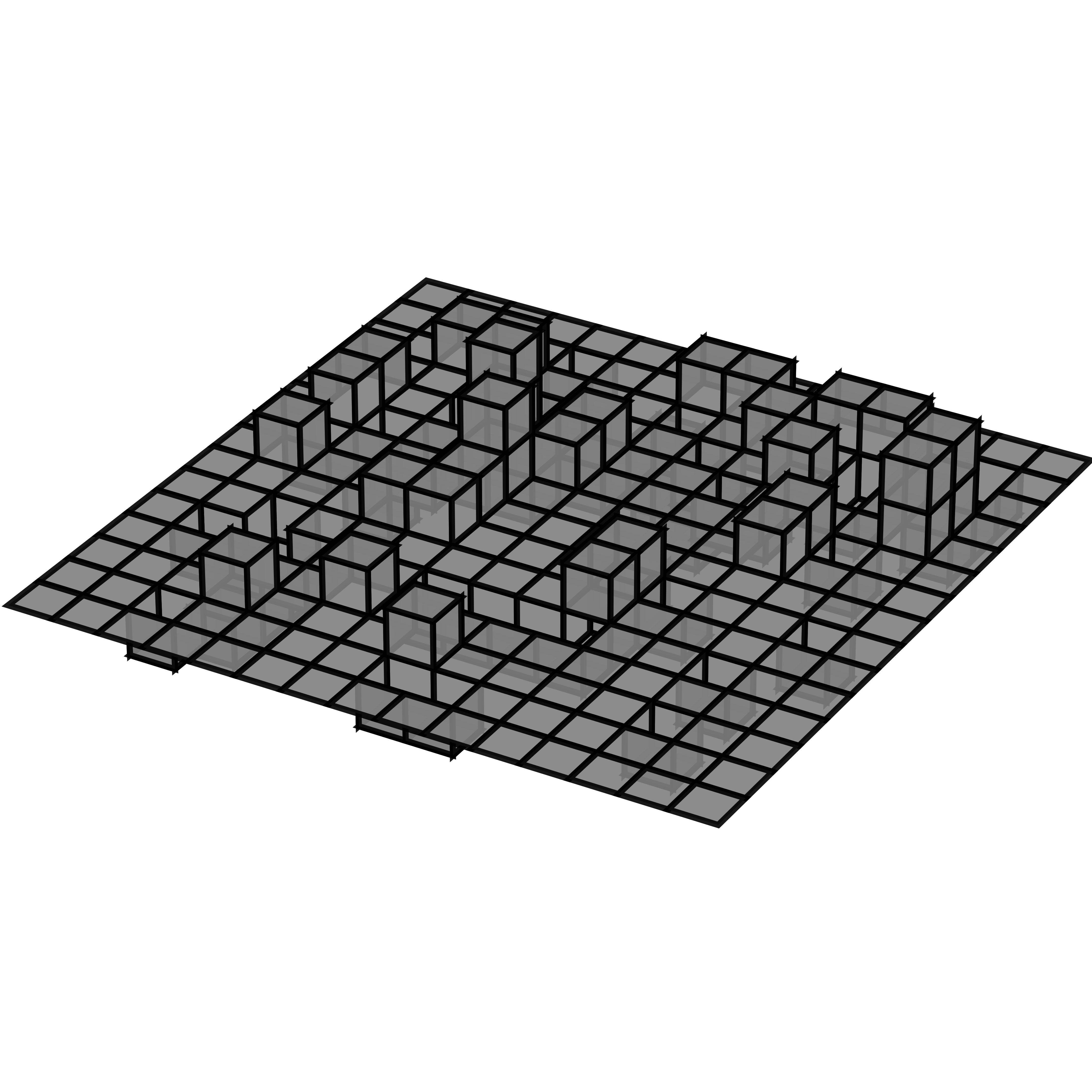}};

        \node[anchor=south] (graph) at ($(surface.north)+(0,0)$) {\includegraphics[width=0.95\linewidth]{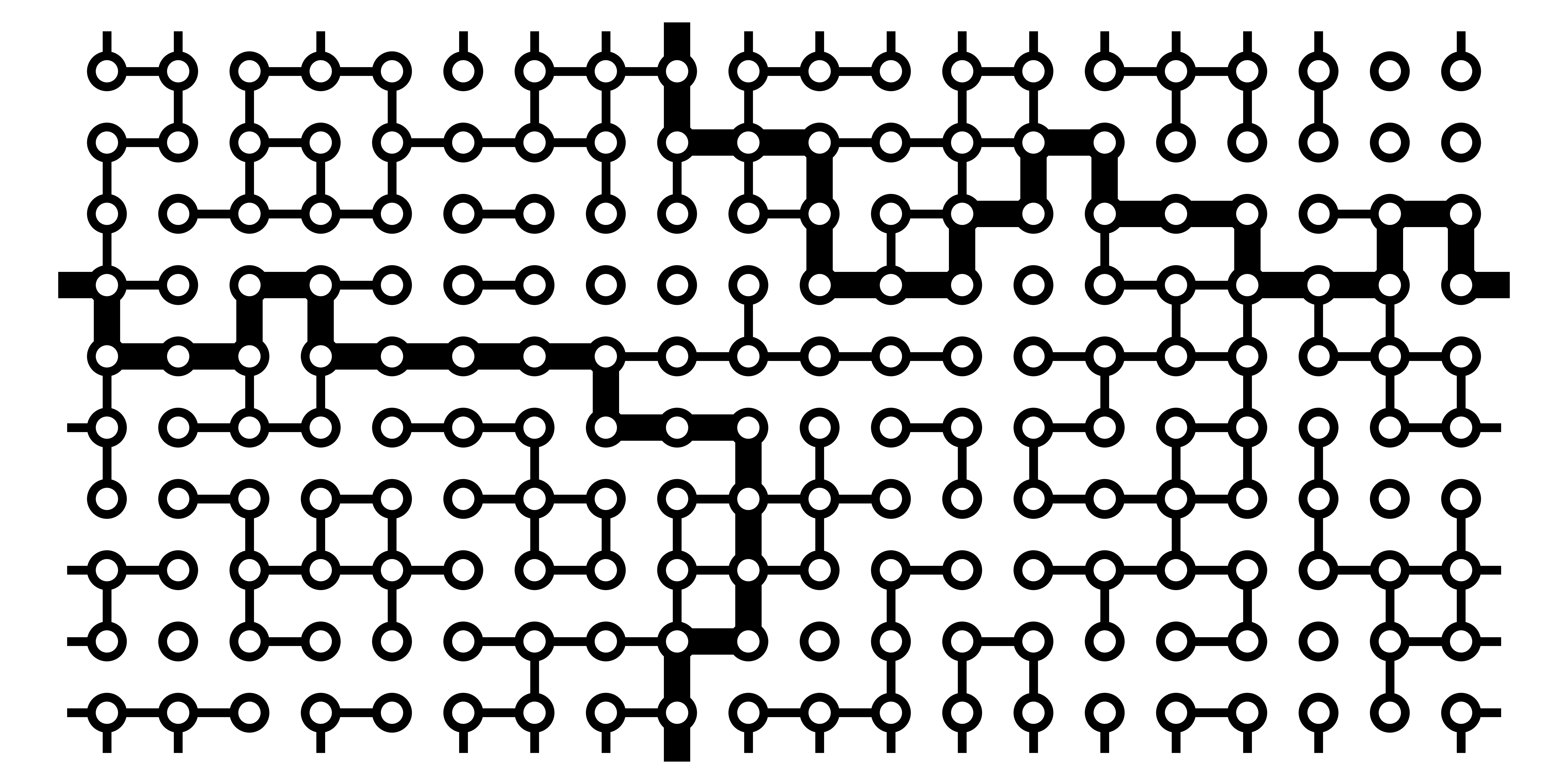}};

    \end{tikzpicture}
    \caption{Homological percolation events: (top) a $1$-dimensional giant cycle is a periodic path; (bottom) a $2$-dimensional giant cycle is a surface spanning the torus. Figures reproduced from~\cite{hom-perc-torus}.}
    \label{figure:giant-cycles} 
\end{figure}

			We can also describe an invasion percolation-type algorithm that relies on \emph{homological percolation}, a generalization of the event that there is periodic path in bond percolation on the torus \cite{prcm-24,hom-perc-euler,hom-perc-giant}. Given a subcomplex $P$ of the torus $\T_N^d,$ homological percolation is the event that $P$ contains a \emph{giant cycle}, an $i$-cycle that ``wraps around'' or ``spans'' the system (in the sense that it is non-trivial in the homology of the torus) \cite{hom-perc-euler}. When $i=1$, the homological percolation event marks the emergence of a periodic path, an event used by Machta et al.~\cite{ic,ic-eq} as a ``topological stopping rule'' for the graphical invaded-cluster algorithm that samples the Potts model on the $2$-torus. Two-dimensional homological percolation occurs when there is a giant (possibly singular) surface of plaquettes spanning the torus in two directions. See Figure~\ref{figure:giant-cycles}.

            
			Motivated by the following result~\cite{prcm-plgt}, the \emph{plaquette invaded-cluster algorithm} (Algorithm \ref{algorithm:pic}) generalizes the invaded-cluster algorithm in \cite{ic} to PLGT.
			
			\vspace{0.25em}
            \begin{theorem}[Duncan and Schweinhart]\label{theorem:ds-8}
                Let $P$ be distributed as the $d$-dimensional PRCM on $\T^{2d}_N$ with parameters $p$ and $q\geq 1$, and $\F$ a field of characteristic not equal to $2$. If $R$ and $S$ are the events that $P$ contains no non-trivial giant cycles in $H_{d}\paren{\T_N^{2d};\F}$\footnote{$\F$ does not need to match $\Z(q)$.} and that $P$ contains giant cycles that form a basis for $H_{d}\paren{\T_N^{2d};\F},$ respectively, then \vspace{-0.4em} 
				\begin{align*}
					\mu(R) \longrightarrow 0 &\hspace{0.5in}p < \sd \\
					\mu(S) \longrightarrow 1 &\hspace{0.5in}p > \sd \\[-2em]
				\end{align*} as $N \longrightarrow \infty,$ where $\sd = \sqrt{q}/(1 + \sqrt{q}).$ 
			\end{theorem}
			
            \noindent Here, $\sd$ corresponds to the \emph{self-dual point} of PLGT under the transformation $\beta \mapsto 1-e^{-\beta}$, conjectured to to be the critical point for Potts lattice gauge theory in four dimensions \cite{kogut}. Moreover, for any $q$, the homological percolation threhsold for the PRCM on $\Z^3$ coincides with the area law--perimeter law transition for Wilson loop variables in the corresponding PLGT (assuming a widely-believed conjecture on the regularity of the supercritical phase of the random-cluster model).\footnote{See Theorem 5 of~\cite{duncan2023sharp} and Theorem 10 of~\cite{prcm-plgt}.}\looseness=-1

           As in Algorithm \ref{algorithm:psw}, the update routine of the invaded-cluster algorithm takes as input a $1$-cochain $f_t \in C^1\paren{X;\Z(q)}$ and passes through a percolation subcomplex $P_{t+1}$ to sample a spin assignment at time $t+1$.

\begin{boxedalgorithm}{Plaquette invaded-cluster (PIC)}{algorithm:pic}


	\begin{enumalgorithm}
		\item \label{algorithm:pic:step:1} Let $(x_n)$ be a uniform random shuffling of the plaquettes of $X$.
		\item Set $k \coloneqq 1$, and let the subcomplex $P_0$ comprise all vertices and edges of $X$.
		\item For each $0 < k \leq n$,
			\begin{enumalgorithm}
				\item\label{algorithm:pic:filtration} add $x_k$ to $P_{k-1}$ if $\delta^1 f(x_k) = 1$, and call the new subcomplex $P_k$;
				\item \label{algorithm:pic:step:check} if $P_k$ has some desired number of giant cycles, set $P_{t+1} \coloneqq P_k$ and go to \ref{algorithm:pic:step:report}.
			\end{enumalgorithm}
		\item \label{algorithm:pic:step:report} Uniformly sample $f_{t+1}$ from $Z^1(P_{t+1};\Z(q))$, and set $t \coloneqq t+1$.
        \item Return to \ref{algorithm:pic:step:1} unless a suitable convergence threshold is met.
	\end{enumalgorithm}
\end{boxedalgorithm}

			\noindent Like the invaded-cluster algorithm for the Potts model, we do not have a rigorous proof that this algorithm targets a distribution that converges to Potts lattice gauge theory as $N\to\infty$. However, the same heuristic described in~\cite{ic} suggests that this is the case.

    \section{Sampling via duality}

        \label{section:duality}
        \noindent Potts lattice gauge theory is dual to the Potts model on $\Z^3$ and is self-dual on $\Z^4$, relationships classically derived by manipulating partition functions via Fourier expansion~\cite{kogut,wegner}. While this interplay lets us describe a model's thermodynamics in terms of its dual, it is not immediately clear how to re-wire an algorithm that (e.g.) samples the Potts model on $\Z^3$ to instead sample its dual Potts lattice gauge theory. The duality transformation for the plaquette random-cluster model, however, allows sampling via duality as it is defined explicitly on the probability space.
    	
    
	    The \emph{dual lattice} $\paren{\Z^d}^{\bullet}$ translates every cell of $\Z^d$ by $\nicefrac 12$ in each coordinate, so each $j$-cell $x \in \Z^d$ intersects exactly one $(d-j)$-cell $x^\bullet \in (\Z^d)^\bullet$ called its \emph{dual cell}. The \emph{dual subcomplex} $X^\bullet \subset (\Z^d)^\bullet$ of an $i$-dimensional subcomplex $X \subset \Z^d$ then comprises dual cells $x^{\bullet} \in X^{\bullet}$ if and only if $x \notin X$. If $X$ is distributed as the PRCM with parameters $p$ and $q$, then $X^{\bullet}$ is distributed as the PRCM with parameters \[ \allowdisplaybreaks p^*\paren{p,q} \coloneqq \frac{(1-p)q}{(1-p)q + p}\] and $q$.\footnote{This was first observed in~\cite{duncan2023sharp}; see Theorem 14 of~\cite{prcm-24} for the most general statement.} To make this correspondence precise, we need to take boundary conditions into account: free boundary conditions in the plaquette random-cluster model correspond to wired boundary conditions for the dual~\cite{prcm-24}. The dual of the PRCM with periodic boundary conditions (that is, on $\T_N^d$) is ``almost'' a PRCM with periodic boundary conditions, but needs to be reweighted by a term counting the number of giant cycles in $P$ to obtain an exact duality.\footnote{See the discussion before Theorem 18 of~\cite{prcm-plgt}.} 
	
	    Equipped with a duality transformation, we can now sample Potts lattice gauge theory with parameters $\beta$ and $q$ on (say) a cube in $\Z^3$. First, sample the Potts model $g$ on the cube with parameters $\beta^*\paren{\beta,q}=\log\paren{\nicefrac{e^{\beta}+q-1}{e^{\beta}-1}}$ and $q$ (under appropriate boundary conditions) via any known technique. Next, let $P^{\bullet}$ be the graph containing each edge $e$ of $(\Z^d)^\bullet$, where $\delta^0 g\paren{e}=1$ independently with probability $p^*\paren{p,q}$. This way, $P^{\bullet}$ is distributed as the classical random-cluster model, so its dual $P$ is distributed as the PRCM. Finally, select a cocyle $f \in Z^1\paren{P;\Z(q)}$ at uniform random to sample PLGT.

\begin{figure*}[t]
	\centering
	\begin{tikzpicture}
		\def\fulltimeBottomMargin{0.2}
		\def\maxBoxWidth{0.03in}
		\def\maxBoxHeight{1.2in}
		\def\topImageWidth{0.46\linewidth}
		\def\bottomImageWidth{0.46\linewidth}
		\def\imageSep{0.1in}
		
		\tikzset{
			blocklabel/.style={font={\footnotesize}, inner sep=3pt},
			toplabel/.style={font={\footnotesize\color{gray}}, inner sep=1pt},
			bottomlabel/.style={font={\scriptsize}, inner sep=2pt},
			padded/.style={inner sep=0pt, draw, ultra thick},
			bounding/.style={shape=rectangle, draw, very thick, fill=none, anchor=north west, minimum width=\maxBoxWidth, minimum height=\maxBoxHeight},
			connection/.style={very thick},
			legend/.style={minimum width=0.25in, minimum height=0.025in},
		};
		
		\node at (0,0) (2) {\includegraphics[width=\topImageWidth]{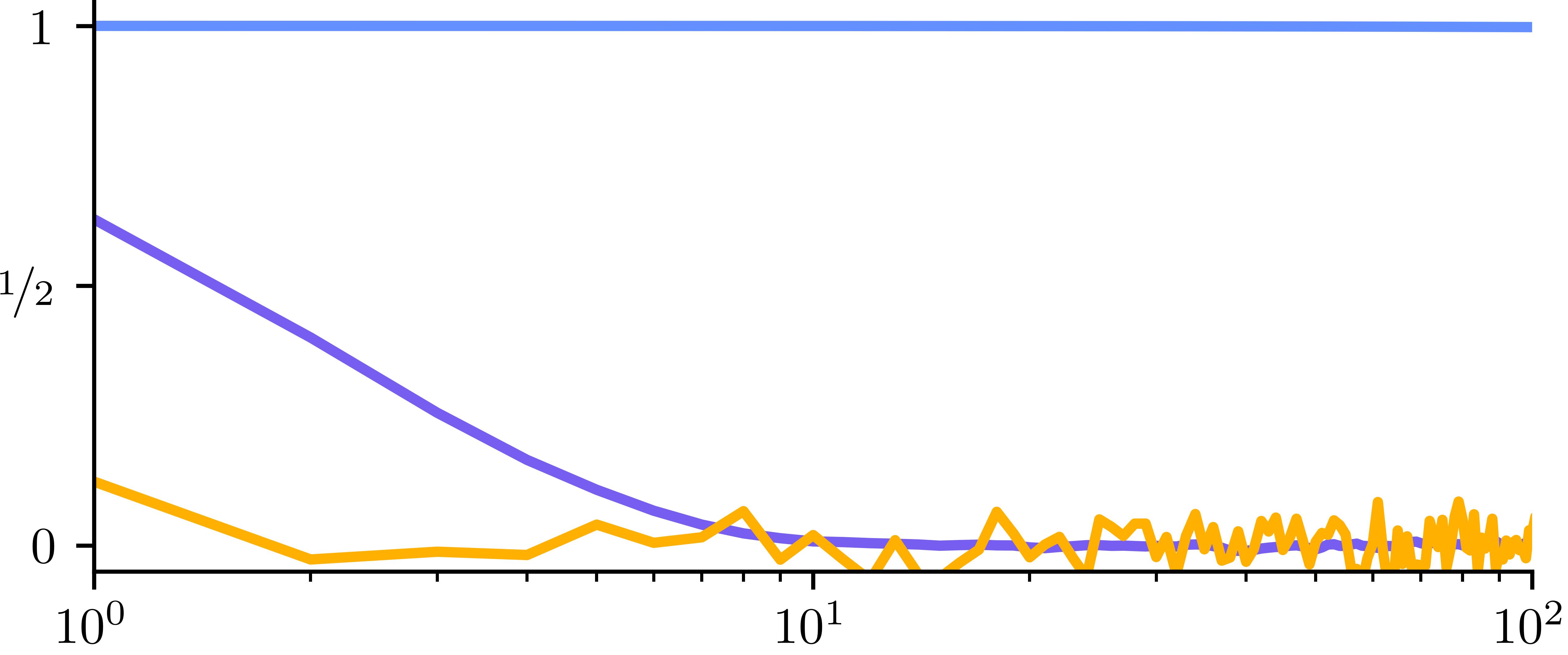}};
		\node[anchor=west] (3) at ($(2.east)+(\imageSep,0)$) {\includegraphics[width=\topImageWidth]{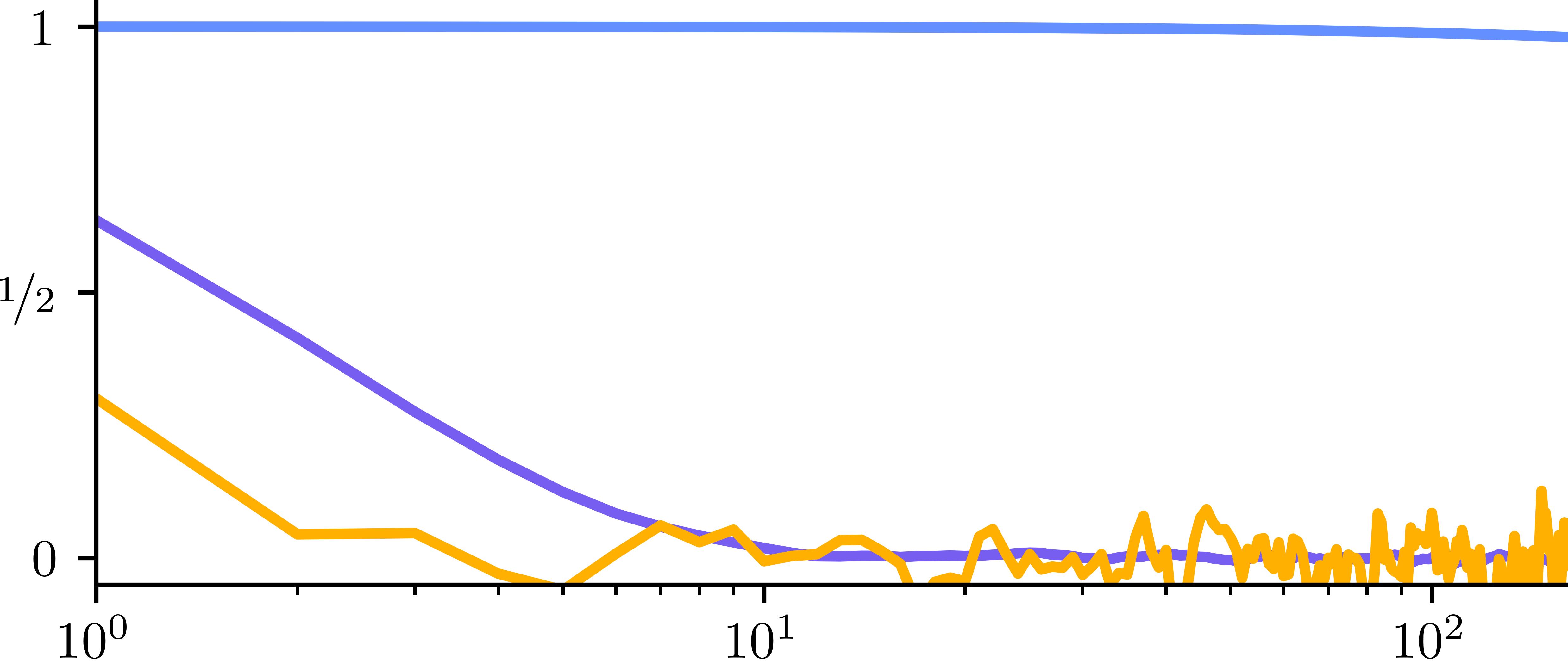}};
		
		\node[toplabel,rotate=90,anchor=south] at (2.west) {autocorrelation};
		\node[toplabel, anchor=north] at (2.south) {($\log_{10}$) iterations};
		
		\foreach \k in {2,3} {
			\node[blocklabel,anchor=south east] at (\k.north east) {$q=\k$};
		}

		\matrix[matrix of nodes, column sep=0.15in, anchor=north] (labels) at ($(2.south east)+(0.05in,-0.125in)$) {
			\node[legend, fill=cornflowerblue] (glauber) {}; \node[blocklabel, anchor=west] at (glauber.east) {plaquette Glauber}; &%
			\node[legend, fill=mediumslateblue] (sw) {}; \node [blocklabel, anchor=west] at (sw.east) {plaquette Swendsen--Wang}; &%
			\node[legend, fill=saffron] (ic) {}; \node [blocklabel, anchor=west] at (ic.east) {plaquette invaded-cluster};\\
		};
	\end{tikzpicture}
	\vspace{-0.5em}
	\caption{Comparing normalized autocorrelation of total energy $\Energy$ for the plaquette Glauber, Swendsen--Wang, and invaded-cluster algorithms on the cubical four-torus $\T^4_{11}$. See Appendix \ref{appendix:exponents} for definitions.}
	\label{figure:sw-autocorrelation}
	\vspace{-1.5em}
\end{figure*}

	    Does the Swendsen--Wang algorithm exhibit the same dynamics as its corresponding procedure on the dual complex? To see that it does not, suppose $P_0 \subset \T_N^{3}$ contains every $2$-cell: any uniform cocycle $f_1\in Z^1\paren{P_0;\Z(q)}$ then vanishes on every $2$-cell of the torus, so $P_1$ is distributed as independent plaquette percolation with probability $p$. On the other hand, $Q_0\coloneqq P_0^{\bullet}$ is the empty graph containing every vertex of $\paren{\T_N^{3}}^\bullet$ but none of its edges. As such, $Z^0\paren{Q_0;\Z(q)}=C^0\paren{Q_0;\Z(q)}$, and a uniform cocycle $g_0$ on $Q_0$ assigns a uniformly random spin in $\Z(q)$ to each vertex of $\paren{\T_N^{3}}^\bullet$. After performing one step of Swendsen--Wang dynamics, the probability that a given edge is contained in $Q_1$ is $\nicefrac{p^*\paren{p,q}}{q}$, and the probability that a dual plaquette is included in $Q_1^{\bullet}$ is $1-\nicefrac{p^*\paren{p,q}}{q}\neq p.$ Moreover, the states of the plaquettes in $Q_1^{\bullet}$ are not independent.

        Sampling the classical random-cluster model on $\Z^3$ via the Swendsen--Wang dynamics on the dual plaquette random-cluster model turns out to be equivalent to sampling from the Loop--Cluster coupling defined in~\cite{zhang2020loop}. We further discuss this fact, and an extension to the generalized $i$-dimensional PRCM (where sampling via duality recovers the generalized Loop--Cluster coupling defined in \cite{hansen2025general}), in Appendix~\ref{appendix:extended-probability}.

    \section{Computation and experiments}
    	\label{section:experiments}

\begin{figure*}[b]
	\centering
	\vspace{-1em}
	
	\def\scale{11}
	
	\begin{tikzpicture}
		\def\imageWidth{0.46\linewidth}
		\def\imageSep{0.16}
		\tikzset{
			toplabel/.style={font={\footnotesize}, inner sep=1pt},
			bottomlabel/.style={font={\scriptsize}, inner sep=2pt},
		};
		
		\node[anchor=east, inner sep=0.05in] (2) at (0,0) {\includegraphics[width=\imageWidth]{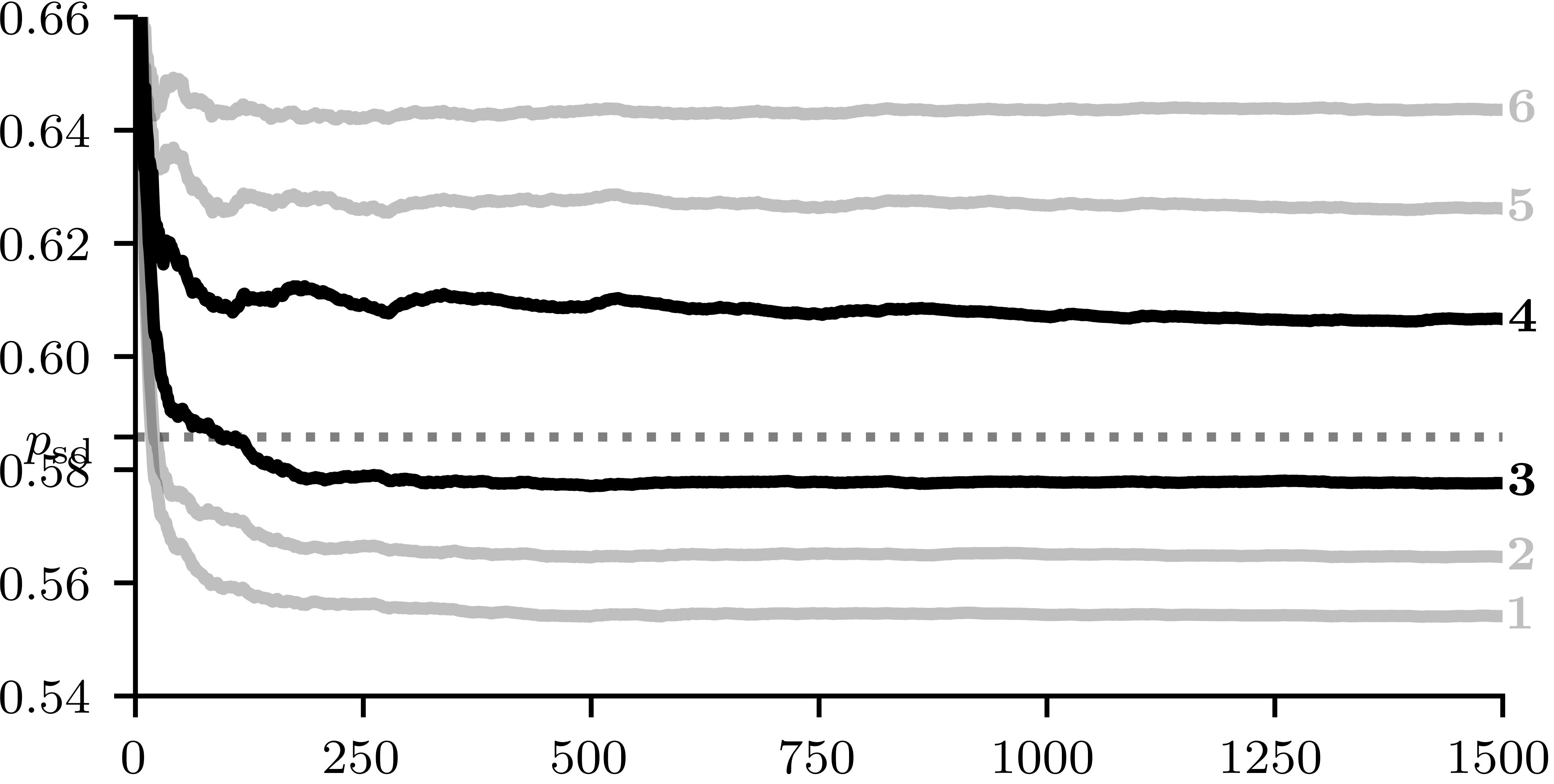}};
		\node[anchor=west, inner sep=0.05in] (3) at (0,0) {\includegraphics[width=\imageWidth]{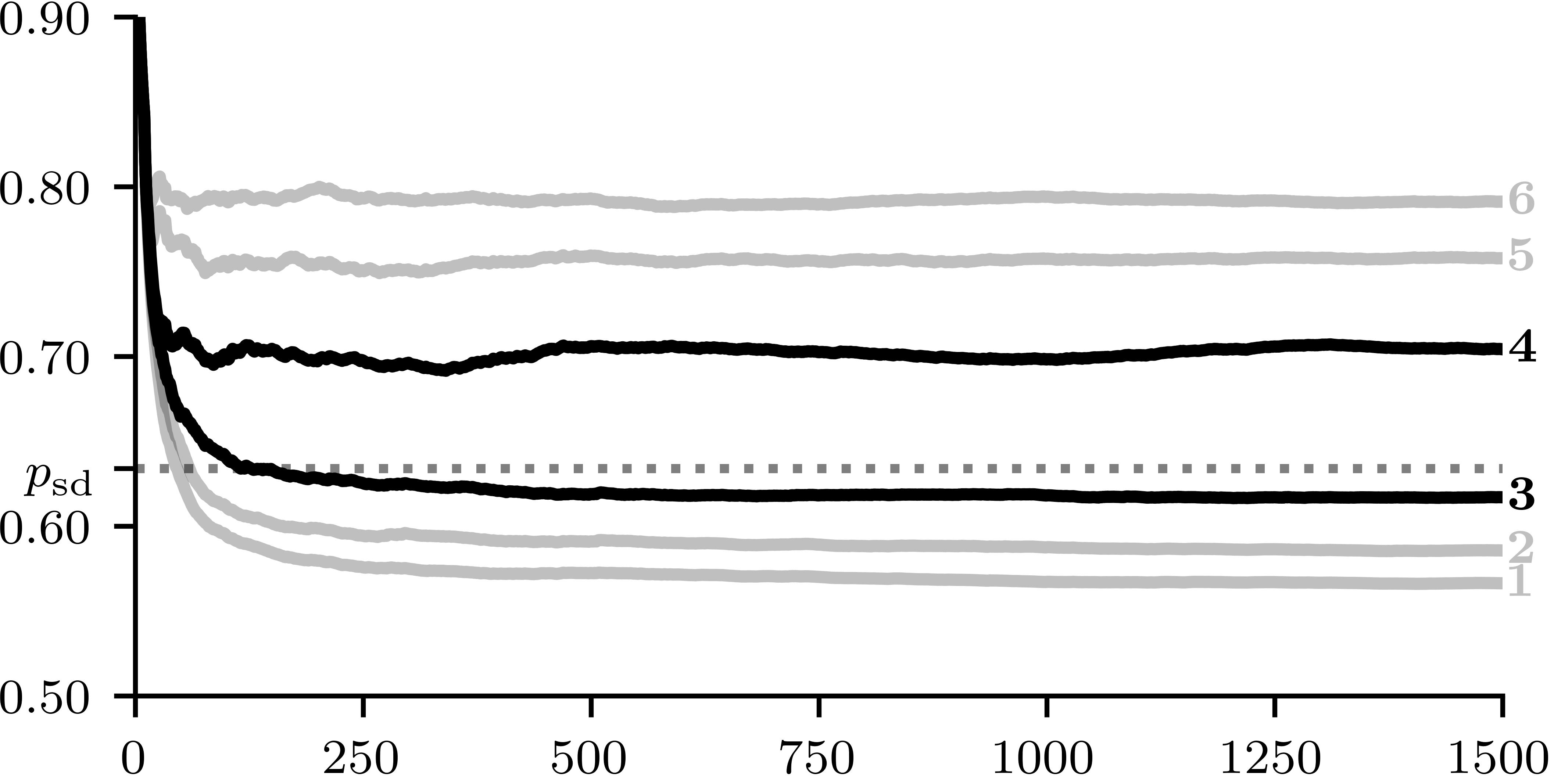}};
		
		\foreach \k in {2,...,3} {
			\node[anchor=north east, toplabel] at ($(\k.north east)-(0.09in,0)$) {$q=\k$};
		}
		
		\node[bottomlabel, anchor=north] at ($(2.south)+(6pt,2pt)$) {\textcolor{gray}{iterations}};
		
		\node[bottomlabel, rotate=90, anchor=south] at ($(2.west)+(0.5pt,3pt)$) {\textcolor{gray}{prop. occupied}};
	\end{tikzpicture}
	\vspace{-0.5em}
	\caption{Proportions of non-frustrated plaquettes included in $P_{t+1}$ (\emph{occupancy}) plotted against iterations of the plaquette invaded-cluster algorithm on the cubical $4$-torus $\T^4_{\scale}$ for $q=2$ (left) and $q=3$ (right) at $\sd$. Each curve is annotated by the number of linearly independent giant cycles encountered. Dark curves indicate number of giant cycles found before re-sampling $f_t$; dotted line indicates the theoretical critical probability $\sd$. At each iteration, we flip a fair coin to decide whether to re-sample $f_t$ after the third or fourth giant cycle is encountered. We will continue to update this figure with data from larger simulations.}
	\label{figure:occupation-rates}
\end{figure*}
    	
    	\subsection{Programmatic implementation}
    		\label{subsection:programming}
    		
     		\noindent From here on out, we restrict attention to the case where $q$ is prime and replace the multiplicative group $\Z\paren{q}$ with the additive group $\Z_q$ of integers modulo $q.$ This allows us to make use of the field structure on $\Z_q$. In particular, the (co)chain and (co)homology groups with coefficients in $\Z_q$ are vector spaces and the coboundary operator $\delta$ is a linear transformation, so we can implement the algorithms using linear algebra. Abusing notation, the \emph{$i$\th coboundary matrix} $\delta^i$ has a row for each $i$-cell $Y$ and a column for each $(i-1)$ cell $y$; the corresponding entries in $\delta^i$ are \[ = 
     			\begin{cases}
     				1_{\Z_q} & y \in \partial Y \\
     				-1_{\Z_q} & \inverse y \in \partial Y \\
     				0 & \text{otherwise.}
				\end{cases}
     		\] As an example, the entries of $\delta^1$ for the edges $\{ab\}$ and $\{ac\}$ of the plaquette $x = \{abdc\}$ in Figure \ref{figure:edge-labeled-cubical-complex} are $1$ and $-1$, respectively.
     		
     		With this setup, sampling cocycles $f$ from $C^1(P;\Z_q)$ becomes an exercise in matrix reduction.\footnote{These matrices are, even for relatively small $N$, extremely large and sparse: the matrix $\delta^1$ of $\T^4_{10}$ has $2.4 \times 10^9$ entries --- of which $2.4 \times 10^5$ are nonzero --- for a density of $0.01\%$.} We use sparse row reduction to solve for a uniform random solution to $\delta^1f = 0$ in step \ref{algorithm:psw:step:sample} of Algorithm \ref{algorithm:psw} and step \ref{algorithm:pic:step:report} of Algorithm \ref{algorithm:pic}.\footnote{Since $\Z_q$ is the additive group of integers mod $q$ in this context, a solution to the system of linear equations $\delta^1 f = 0$ gives us a cochain $f$ whose entries in $\Z(q) \subset \C$ are a cocycle on $P$ as defined in Section~\ref{section:introduction:subsection:PLGT}.}
			
			To determine whether we have homologically percolated sufficiently many times --- that is, to verify the condition in step \ref{algorithm:pic:step:check} of Algorithm \ref{algorithm:pic} --- we use \emph{persistent homology}, which identifies topological features of $P$ that are also topological features of $X$. We first break $X$ into a nested sequence of subcomplexes called a \emph{filtration}: the first $t$ elements of the filtration are the subcomplexes $P_t$ specified by Algorithm \ref{algorithm:pic}, and the remaining elements are the subcomplexes formed by adding each plaquette \emph{not} in $P$ to $P_t$ one at a time.
            
            At each step of the filtration, we then ask whether the added plaquette creates a new cycle or annihilates an existing one --- the cycles created at or before time $t$ minus the cycles annihilated after time $t$ are the giant cycles in $P$, and form the \emph{first persistent homology group} $PH_1(P)$. Computing (the rank of) this group amounts to re-ordering the columns of the boundary matrices, then reducing to \emph{Smith normal form} \cite{comptop,zomorodian}. Algorithms \ref{algorithm:psw} and \ref{algorithm:pic}, alongside supporting arithmetic and topological routines, are implemented in \emph{Algebraic Topology-Enabled AlgorithMs for Spin systems} (\ATEAMS), this work's companion software library \cite{ateams}; for more, see Appendix \ref{appendix:ATEAMS}.

    	\subsection{Experiments and results}
    		\label{subsection:experimental-results}
    		
			\noindent Experimentally, we restrict our attention to $2$-dimensional plaquette percolation on the $4$-dimensional torus with coefficients in $\Z_q$ for prime $q$. In these systems, the critical point is conjectured to be the self-dual point $\sd = \sqrt{q}/(1 + \sqrt{q})$ \cite{prcm-plgt}. Using our implementations of the plaquette Swendsen--Wang and invaded-cluster algorithms, we compare their performance to each other and to single-spin flip (or \emph{Glauber}) dynamics on $\T^4_{16}$, the cubical $4$-torus of scale $N=16$.
			
			We first compare the \emph{normalized autocorrelation decay} of each algorithm, tracking the statistical similarity between total energies $\H(f_0)$ and $\H(f_t)$ as $t$ grows. These data, shown in Figure \ref{figure:sw-autocorrelation}, clearly reflect the algorithms' sampling methodologies: the destructive Swendsen--Wang and invaded-cluster algorithms lose memory of the initial cochain $f_0$ within ten iterations. Glauber dynamics are eidetic, remembering $f_0$ for most of the run-time.
            
            It is important to emphasize that the time scale in Figure \ref{figure:sw-autocorrelation} corresponds to \emph{iterations} of each algorithm, not the clock time it took to complete them. Plaquette Glauber dynamics bears a comparatively light computational load, while the Swendsen--Wang and invaded-cluster algorithms require more time and resources. As Glauber dynamics edits a single entry of the input cochain, it is only memory-constrained by the scale and dimensionality of the ambient lattice. The plaquette Swendsen--Wang and invaded-cluster algorithms, on the other hand, need to solve a large, sparse system of linear equations at each iteration, and so are time- \emph{and} memory-bound by the lattice parameters. Finally, the invaded-cluster algorithm's behavior hews closely to Swendsen--Wang's at criticality. This differs from the observed behavior of these algorithms for the Potts model, but may be an artifact of the system size. Performance statistics for each model are included in Appendix \ref{appendix:ATEAMS:performance}.
			
			Second, we consider the effect of the topological stopping rule for the plaquette invaded-cluster algorithm. By duality, the probability that the PRCM (sampled at the self-dual point) has at least one giant cycle exceeds $\nicefrac 12$. As a consequence of Theorem~\ref{theorem:ds-8}, we know that as $N\to\infty$ the PRCM has --- with high probability --- no giant cycles when $p<\sd$, and a full linearly independent set of $\binom{4}{2}=6$ giant cycles when $p>\sd$. However, we have no \emph{a priori} prediction for the distribution of the number of giant cycles at $\sd$ besides knowing that it is symmetric and centered about $3$. Figure \ref{figure:occupation-rates} shows the mean percentage of non-frustrated plaquettes included in $P_{t+1}$ before encountering the $j$\th linearly independent giant cycle (for $1 \leq j \leq 6$). This quantity is closest to $\sd$ for $j=3$ and, as the scale of the torus is taken to infinity, we expect all of these curves to converge to $\sd$. We conjecture further that the local statistics of sampled plaquette percolations (and coupled spin assignments) will approach those of the self-dual PRCM (and PLGT) for any stopping rule, but a gap remains for $N=11$.

		

\begin{table}[t]
    \centering
    \begin{tabular}{c|cc}
$N$ & $\tau_{\mathcal N,2}$ & $\tau_{\mathcal E,2}$ \\ \hline
$6$ & $1.5145 \pm 0.0345$ & $1.7230 \pm 0.0422$ \\
$8$ & $1.5516 \pm 0.0361$ & $1.7331 \pm 0.0425$ \\
$11$ & $1.5678 \pm 0.0364$ & $1.7604 \pm 0.0431$ \\
$16$ & $1.6359 \pm 0.0394$ & $1.8591 \pm 0.0470$ \\
\end{tabular}

    \vspace{-0.5em}
    \caption{Estimated integrated autocorrelation times for the PSW algorithm with $q=2$.}
    \label{table:psw:2}

    \vspace{1em}
    
    \begin{tabular}{c|cc}
$N$ & $\tau_{\mathcal N,3}$ & $\tau_{\mathcal E,3}$ \\ \hline
$4$ & $1.6688 \pm 0.0402$ & $1.7990 \pm 0.0448$ \\
$6$ & $1.6699 \pm 0.0402$ & $1.8146 \pm 0.0452$ \\
$8$ & $1.6892 \pm 0.0407$ & $1.8352 \pm 0.0457$ \\
$11$ & $1.6945 \pm 0.0408$ & $1.8357 \pm 0.0465$ \\
$16$ & $1.7371 \pm 0.0426$ & $1.8691 \pm 0.0473$ \\
\end{tabular}

    \vspace{-0.5em}
    \caption{Estimated integrated autocorrelation times for the PSW algorithm with $q=3$.}
    \label{table:psw:3}
    \vspace{0.5em}
\end{table}

\begin{figure}[t]
	\centering
	
	\def\field{3}
	
	\begin{tikzpicture}
		\tikzset{
			label/.style={font={\footnotesize\color{gray}}, inner sep=0pt, fill=none}
		}
		\def\imgwidth{0.93\linewidth}
		\def\imgsep{0.0825in}
		\def\xlabeladj{7pt}
		
		\node (energy) at (0,0) {\includegraphics[width=\imgwidth]{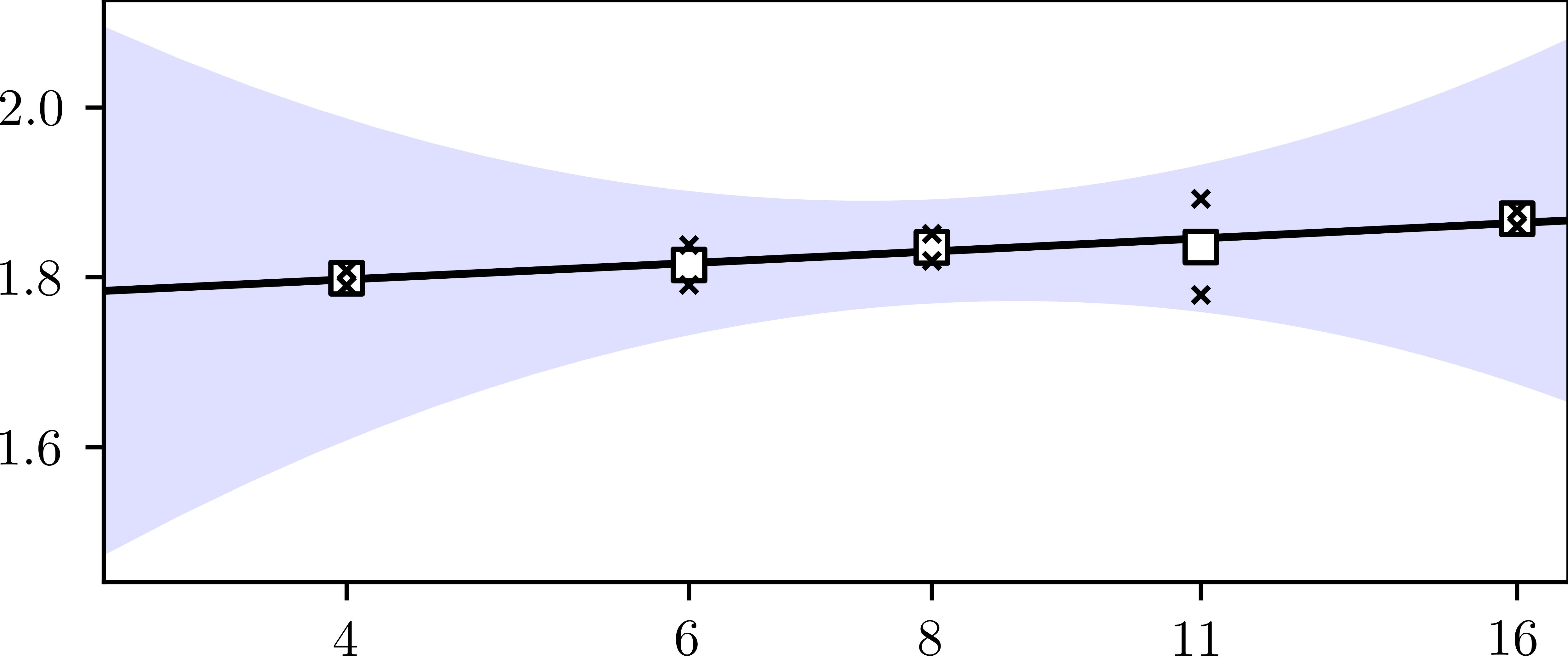}};
		\node[anchor=north] (occupancy) at ($(energy.south)+(0,-\imgsep)$) {\includegraphics[width=\imgwidth]{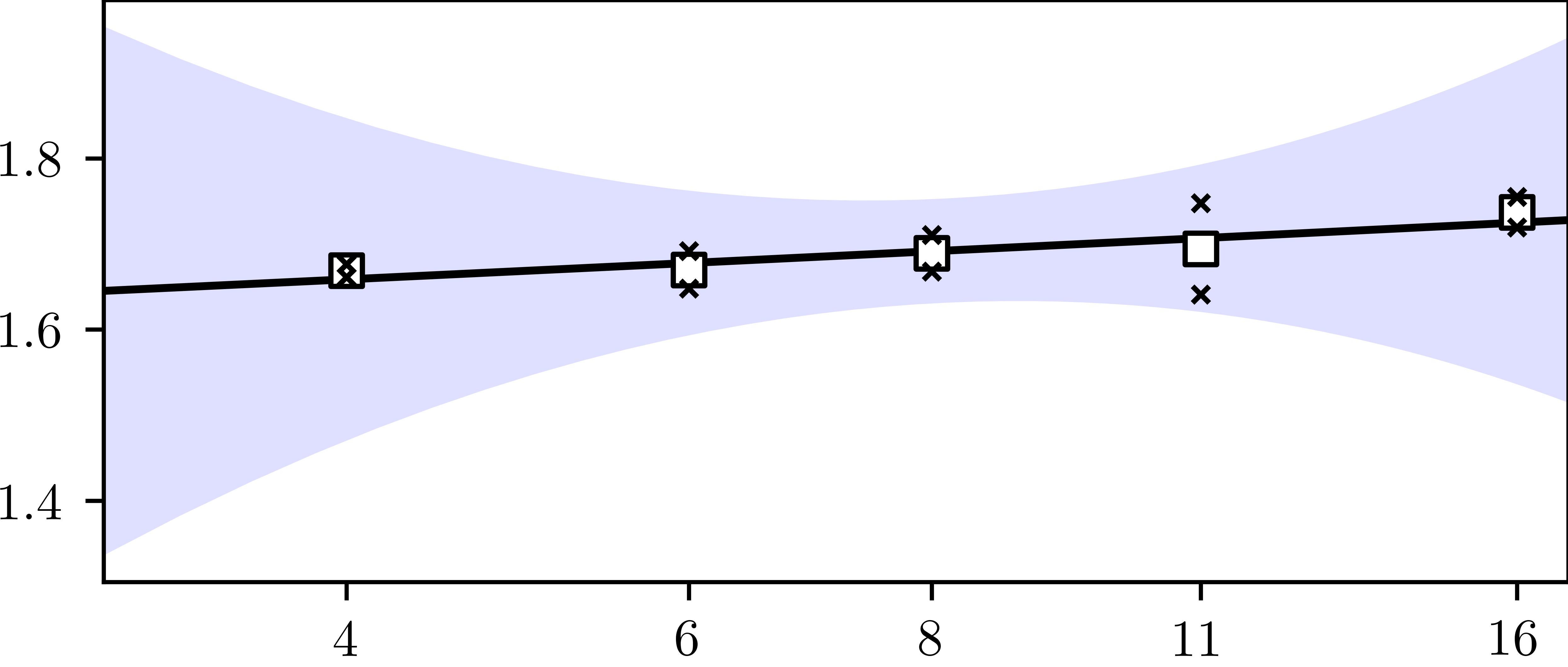}};
		
		\node[label] at ($(energy.south)+(\xlabeladj,0)$) {$N$};
		\node[label,rotate=90,anchor=south west] at (energy.west) {$\tau_{\Energy,\field}$};
		
		\node[label,rotate=90,anchor=south west] at (occupancy.west) {$\tau_{\Occupancy,\field}$};
	\end{tikzpicture}
	\vspace{-1em}
	\caption{Power law fits for $\tau_{\Energy,\field}$ (top) and $\tau_{\Occupancy,\field}$ (bottom). Shaded regions are $68\%$ (one standard deviation) confidence intervals for the fit. Square markers ($\square$) indicate amortized autocorrelation times; cross markers ($\times$) indicate actual.}
	\label{figure:dynamical-exponents}
\end{figure}

			Third, we estimate the \emph{integrated autocorrelation times} $\tau_{\Energy,q}$ and $\tau_{\Occupancy,q}$ for the total energy $\Energy$ and occupancy $\Occupancy$ at varying lattice scales $N$ and $q \in \{2,3\}$, then use these times to estimate the observables' \emph{dynamical critical exponents}. Largely adhering to the methodology in \cite{ossola-sokal}, we perform two critical-temperature PSW runs of length $10^6$, discard the first $9 \times 10^5$ samples, then compute the integrated autocorrelation times with the remainder. Tables~\ref{table:psw:2} and~\ref{table:psw:3} convey our numerical estimates for $\tau_{\Energy,q}$ and $\tau_{\Occupancy,q}$ with error bars, giving a $68\%$ confidence interval for the true value.\footnote{For an observable $Z$, the quantity $\tau_{Z,q}$ is computed by summing normalized (auto)covariances $\Cov(Z_i, Z_{i+t})$ at time lags $1 \leq t \leq M \ll n$, for $n$ the sample size of $Z$. These covariances have a(n unwieldy, but) known expression, from which we can derive the exact value of, as well as a far more convenient approximation for, $\Var(\tau_Z)$. For more detail, see Appendix~\ref{appendix:exponents}.} We then fit the autocorrelation times to a \emph{power law} $A \cdot N^z$, where $z$ is the \emph{critical exponent} to be estimated. Figure \ref{figure:dynamical-exponents} shows fitted data for $\tau_{\Energy,3}$ and $\tau_{\Occupancy,3}$. In the $q=2$ case,\footnote{We omit $\tau_{-,2}$ at the $N=4$ level, as both estimates were extreme outliers.} our estimates are \[z_{\Energy,2} = 0.078 \pm 0.021 \qand z_{\Occupancy,2} = 0.075 \pm 0.011 \] and, for the $q=3$ case, \[z_{\Energy,3} = 0.026 \pm 0.004 \qand z_{\Occupancy,3} = 0.028 \pm 0.007, \] where the quantities $\pm~\varepsilon$ are one standard deviation parameter estimation errors from the power law fit; these results are consistent with $z_{\Energy,q} \approx z_{\Occupancy,q}$ for the plaquette Swendsen--Wang algorithm sampling $q$-state PLGT. The corresponding equality for the Swendsen--Wang algorithm sampling the $q$-state Potts model is an ``almost-theorem'' from~\cite{salasDynamic1997}: the Li--Sokal bound~\cite{liRigorous1989} asserts that the integrated autocorrelation times associated to $\Energy$ and $\Occupancy$ satisfy \[ \tau_{\Occupancy,q} \leq \tau_{\Energy,q} \leq \frac{\tau_{\Occupancy,q}}{\overline \rho_{\Occupancy,q}(1)}, \tag{2.28, \cite{salasDynamic1997}} \] where $\overline \rho_{\Occupancy,q}(t)$ is the normalized autocorrelation time at time $t$. The ``almost-theorem'' says that, if we adopt the mild assumptions that $0 \ll \overline \rho_{\Occupancy,q}(1)$ and $\overline \rho_{\Occupancy,q}(1) \to 1$ as $N \to \infty$, we get equality of the dynamic critical exponents $z_{\Occupancy,q} = z_{\Energy,q}$ at the critical temperature~(2.29, \cite{salasDynamic1997}). See Appendix~\ref{appendix:exponents} for further detail.

    \section{Discussion and conclusions}
    	\label{section:discussion-conclusions}
    	
		\noindent Using tools borrowed from algebraic topology, we describe versions of the \emph{Swendsen--Wang} and \emph{invaded-cluster} algorithms for targeting \emph{$q$-state Potts lattice gauge theory} and the \emph{plaquette random-cluster measure} on cubical tori of arbitrary dimension \cite{prcm-plgt, hom-perc-giant, ic}. These generalized procedures --- the \emph{plaquette Swendsen--Wang} and \emph{plaquette invaded-cluster} algorithms (Algorithms \ref{algorithm:psw} and \ref{algorithm:pic}) --- rely on basic linear algebra and probability. We prove that the plaquette Swendsen--Wang algorithm targets Potts lattice gauge theory, and conjecture that the stationary distribution of the plaquette invaded-cluster algorithm converges weakly to Potts lattice gauge theory in the infinite-volume limit. 
        Further, we empirically demonstrate that the plaquette invaded-cluster algorithm obeys the scaling behavior theorized in~\cite{prcm-plgt}. Finally, we estimate some \emph{dynamical critical exponents} of the plaquette Swendsen--Wang algorithm, and show that their behavior is similar to that previously observed for the classical Swendsen--Wang algorithm.
		
		Each algorithm is implemented in \ATEAMS \cite{ateams}. We detail the memory-efficient data structures and  subroutines underlying the computational simulations, and believe the implementation to be a robust experimental tool for studying spin systems. In future work, we plan to: further optimize \ATEAMS; gather experimental data regarding the scaling behavior of PLGT and PRCM at criticality; estimate the dynamical critical exponents of the invaded-cluster algorithm; extend these algorithms to simulate the \emph{Potts lattice Higgs} and \emph{coupled plaquette percolation} models~\cite{eldridgeCellular2026a} and build on experimental work in~\cite{somozaSelfDual2021}; use the plaquette invaded-cluster algorithm to estimate homological percolation thresholds in higher-dimensional settings; and extend these methods to other lattice gauge theories.
        
        
        \FloatBarrier
    	
    \section*{Acknowledgements}
    	\noindent Anthony E. Pizzimenti was generously supported by the NSF Graduate Research Fellowship Program (\#2024370283). The authors are grateful to Summer Eldridge for her contributions to (and testing of) \ATEAMS, and to Ramgopal Agrawal, Madeline Horton, and Shrunal Pothagoni for their thoughtful comments on the form and content of this work.
\nocite{creutz1979experiments}

	{
		\printbibliography
	}

    \appendix
	\section{Definitions}
		\subsection{Algebraic topology}
			\label{appendix:definitions}
		    
		    \noindent Let $e_i$ denote the line segment between the origin in $\Z^d$ and the endpoint of the vector $\vec{e}_i$ (for $1 \leq i \leq d$). The set $\D = \prod e_i$ is then a unit \emph{$d$-cube} (or \emph{$d$-cell}). 
		    
		    To convert geometry to algebra, we treat point-set topological boundaries as finite linear combinations of the faces composing them: the set $C_i(X,G)$ is the \emph{$i$\th chain group} of the cubical complex $X$ with coefficients in the abelian group $G$. An \emph{$i$-chain} in $C_i(X,G)$ is a linear combination \[ \gamma = \sum \alpha_j x_j, \] where the $\alpha_j$ are in $G$ and the $x_j$ $i$-cells of $X$. The \emph{$i$\th boundary map} $\partial_i: C_i(X,G) \to C_{i-1}(X,G)$ takes an $i$-cell $x$ to an oriented sum of its faces. Ordering the set $F = (F_1, \dots, F_n)$ of $x$'s faces lexicographically by their vertex coordinates, the boundary of $x$ is \[ \partial_i(x) = \sum_{k=1} (-1)^{k-1} F_k.\] Figure \ref{figure:cubical-complex-ordering} demonstrates $\partial_2$ when applied to a $2$-cube (or \emph{square plaquette}) --- edges are always oriented to ``point'' from the lesser to the greater coordinate, so the boundary of the square is the alternating sum over its edges in lexicographic order.\footnote{The face ordering is arbitrary so long as the $\partial_{i-1}\partial_i=0$ condition is satisfied. We choose lexicographic order for its relative simplicity, ease of computation, and precedent in relevant literature \cite{druhl-wagner}.} Moreover, the boundary maps are homomorphisms: given an $i$-chain $\gamma = \sum \alpha_j x_j$,\vspace{-0.5em}
		        \begin{align*}
		        	\partial_i(\gamma) &= \partial_i\left( \sum \alpha_j x_j \right) \\
		        	&= \sum \alpha_j \partial_i(x_j).
		        \end{align*}
		        These maps also satisfy the relation $\partial_{i-1}\partial_i = 0$ for all $i$ (where $\partial_0 = 0$), as shown in Figure \ref{figure:cubical-complex-ordering}. All told, we have a \emph{chain complex} \vspace{-0.5em}
				\[\begin{tikzcd}[cramped,column sep=small, row sep=tiny]
					{C_i(X,G)} & {C_{i-1}(X,G)} & \cdots & {C_0(X,G)} & 0,
					\arrow["{\partial_i}", from=1-1, to=1-2]
					\arrow["{\partial_{i-1}}", from=1-2, to=1-3]
					\arrow["{\partial_1}", from=1-3, to=1-4]
					\arrow["{\partial_0}", from=1-4, to=1-5]
				\end{tikzcd}\vspace{-0.5em}\] a sequence of abelian groups connected by their respective boundary maps. Because the composition of consecutive boundary maps is zero, we know that $\im(\partial_{i+1}) \subseteq \ker(\partial_i)$ --- where the chains in $\im(\partial_{i+1})$ are called \emph{boundaries} and those in $\ker(\partial_i)$ called \emph{cycles} --- and so define the \emph{$i$\th cubical homology group} with coefficients in $G$ as the quotient group $H_i(X,G) \coloneqq \ker(\partial_i)/\im(\partial_{i+1})$. When $G = \F$ is a field (as is the case in the computational experiments), the $H_i(X, \F)$ are vector spaces over $\F$. Moreover, when $X$ is finite, the boundary operators $\partial_i$ can be represented as matrices with entries in $\F$ \cite{hatcher, comptop, comphom, topillustrated}.
				
				\emph{Cochains} play the dual role to chains: the \emph{$i$\th cochain group} with coefficients in $G$ is $C^i(X;G) \coloneqq \Hom(C_i(X,G),G)$, the group of homomorphisms from the $i$\th chain group into $G$. The \emph{$i$\th coboundary operator} $\delta^i$ takes an $i$-cochain $f$ to an $(i+1)$-cochain by evaluating $f$ on boundaries of $(i+1)$-chains. Given an $(i+1)$-chain $\gamma$, we set \[(\delta^i f)(\gamma) \coloneqq f(\partial_{i+1}(\gamma)),\] so the composition $f \partial_{i+1}$ is an $(i+1)$-cochain. We get a dual \emph{cochain complex} \vspace{-0.5em}
				\[\begin{tikzcd}[cramped,column sep=small]
					{C^{i+1}(X;G)} & {C^i(X;G)} & \cdots & {C^0(X;G)} & 0
					\arrow["{\delta^{i}}"', from=1-2, to=1-1]
					\arrow["{\delta^{i-1}}"', from=1-3, to=1-2]
					\arrow["{\delta^0}"', from=1-4, to=1-3]
					\arrow["0"', from=1-5, to=1-4]
				\end{tikzcd}\vspace{-0.5em}\] connecting the cohomology groups by coboundary maps with the property that $\delta^{i+1}\delta^i = 0$. The \emph{$i$\th cubical cohomology group} with coefficients in $G$ is the quotient group $H^i(X;G) \coloneqq \ker(\delta^{i+1})/\im(\delta^i)$. A consequence of the \emph{universal coefficient theorem for cohomology} tells us that $H_i(X, G) \cong H^i(X; G)$ when $G$ is a field and $H_i(X,G)$ is \emph{finitely generated}. As with boundary operators, the coboundary operators $\delta^i$ can be represented as matrices with entries in $\F$ when $X$ is finite; because the boundary and coboundary operators are exact duals under these conditions, $\delta^{i-1} = \partial_i^\top$ \cite{hatcher, topillustrated}.
				
				\begin{figure}
	\centering
	\vspace{-0.5em}
	\begin{tikzpicture}
		\def\width{0.45}
		\def\cubesep{1.1}
		\def\lowerlabelsep{0.75}
		\def\opacity{0.5}
		\def\offset{0.05in}
		\def\labeloffset{0.05in}
		
		\tikzset{
			closed/.style={ultra thick, dashed},
			open/.style={ultra thick},
			square/.style={shape=rectangle, minimum size=\width in, fill opacity=\opacity, draw=none},
			supertinylabel/.style={font=\tiny, inner sep=2pt},
			label/.style={font=\footnotesize, inner sep=2pt},
			biglabel/.style={font=\footnotesize, inner sep=6pt},
			tinylabel/.style={font=\footnotesize, inner sep=5pt},
			circ/.style={draw,circle,fill=black,inner sep=1pt},
			middy/.style={currarrow,pos=0.5,sloped,scale=0.75}
		}
		
		\node[square, pattern=north west lines] (left) at (0,0) {};
		\draw (left.north west) [open]-- (left.north east) node[middy] {};
		\draw (left.north west) [open]-- (left.south west) node[middy,xscale=-1] {};
		\draw (left.south west) [open]-- (left.south east) node[middy] {};
		\draw (left.south east) [open]-- (left.north east) node[middy] {};
		
		\foreach \anchor in {north west,north east,south west,south east} {
			\node[circ] at (left.\anchor) {};
		}
		
		\node[supertinylabel, anchor=north east] at (left.south west) {$(0{,}0)$};
		\node[supertinylabel, anchor=north west] at (left.south east) {$(0{,}1)$};
		\node[supertinylabel, anchor=south east] at (left.north west) {$(1{,}0)$};
		\node[supertinylabel, anchor=south west] at (left.north east) {$(1{,}1)$};
		
		\node[label, anchor=north] at ($(left.south)+(0,-\lowerlabelsep)$) {$X$};
		
		\node[square] (right) at (\cubesep in,0) {};
		\node[square,minimum size=\width in+\offset] (bigright) at (right.center) {};
		\draw ($(right.north west)+(0,\offset)$) [open]-- ($(right.north east)+(0,\offset)$) node[middy,xscale=-1] {};
		\draw ($(right.north west)+(-\offset,0)$) [open]-- ($(right.south west)+(-\offset,0)$) node[middy] {};
		\draw ($(right.south west)+(0,-\offset)$) [open]-- ($(right.south east)+(0,-\offset)$) node[middy] {};
		\draw ($(right.south east)+(\offset,0)$) [open]-- ($(right.north east)+(\offset,0)$) node[middy] {};
		
		\node[label, anchor=north] at ($(right.south)+(0,-\lowerlabelsep)$) {$\partial_2(X)$};
		
		\foreach \ranchor/\xoffset/\yoffset in {north west/0/\offset,north east/0/\offset, north west/-\offset/0,south west/-\offset/0,south west/0/-\offset,south east/0/-\offset,south east/\offset/0,north east/\offset/0} {
			\node[circ] at ($(right.\ranchor)+(\xoffset,\yoffset)$) {};
		}
		
		\foreach \lanchor/\banchor/\sign in {north/south/+,east/west/-,west/east/+,south/north/-} {
			\node[biglabel,anchor=\lanchor] at (bigright.\banchor) {$\sign$};
		}
		
%
		
		\node[square] (far) at (2*\cubesep in,0) {};
		\node[square,minimum size=\width in+\offset] (bigfar) at (right.center) {};
		
		\foreach \name/\ranchor/\xoffset/\yoffset in {A/north west/0/\offset,B/north east/0/\offset, C/north west/-\offset/0,D/south west/-\offset/0,E/south west/0/-\offset,F/south east/0/-\offset,G/south east/\offset/0,H/north east/\offset/0} {
			\coordinate (\name) at ($(far.\ranchor)+(\xoffset,\yoffset)$);
			\node[circ] at ($(far.\ranchor)+(\xoffset,\yoffset)$) {};
		}
		
		\foreach \coord/\labelanchor/\sign in {A/south/+,B/south/-,C/east/-,D/east/+,E/north/-,F/north/+,G/west/-,H/west/+} {
			 \node[tinylabel,anchor=\labelanchor] at (\coord) {$\sign$};
		}
		
		\node[label, anchor=north] at ($(far.south)+(0,-\lowerlabelsep)$) {$\partial_1(\partial_2(X))$};
	\end{tikzpicture}
	\vspace{-1.5em}
	\caption{For $X$ the unit $2$-cube $e_1 \times e_2$ (left, with vertices labeled): $X$'s bounding $1$-chain (center); the bounding $0$-chain of $X$'s bounding $1$-chain (right).}
	\label{figure:cubical-complex-ordering}
\end{figure}
				
				Given a subcomplex $Q \subseteq X$, inclusion ${\iota: Q \into X}$ induces a map ${\iota^*: H_i(Q) \to H_i(X)}$ on homology. More formally, the map $\iota^*$ takes equivalence classes of $i$-chains in $Q$ to equivalence classes of $i$-chains in $X$, and classes $[\gamma] \in H_1(Q)$ are nonzero in $H_1(X)$ only when $\gamma$ does not bound an $(i+1)$-chain in $Q$ or in $X$. The image of $\iota^*$ is the \emph{$i$\th persistent homology group} ${PH_i(Q) \coloneqq \im(\iota^*)}$ \cite{comptop,zomorodian}. In the parlance of homological percolation, these persistent $i$-cycles are \emph{giant} or \emph{essential} cycles, and are present whenever $\iota^*$ is not the zero map \cite{hom-perc-euler,hom-perc-giant}. For our purposes, $X$ is typically the discrete torus $\T^d_N$ formed by identifying opposite faces of the $d$-cube $[0,N]^d$.

            \subsection{Cohomology example} \label{appendix:cohomological-rank}
                \noindent To demonstrate the computation of $\abs{H^1(P;\Z_q)}$, we do so for the $2$-complex in Figure~\ref{figure:edge-labeled-cubical-complex} under the assumption that $q$ is prime. Intuitively, $P$ appears to have two independent cycles that are not boundaries, so we would guess that cocycles modulo coboundaries has two degrees of freedom. Let's confirm: here, $P$ has $19$ edges so $C^1\paren{P;\Z_q}$ is a $19$-dimensional $\Z_q$-vector space. The requirement that $\delta^1 f$ vanishes on the plaquettes adds four linear constraints, so $Z^1\paren{P;\Z_q}$ is $15$-dimensional. $P$ has $14$ vertices and elements of $C^0\paren{P;\Z_q}$ give rise to the same element of $B^1\paren{P;\Z_q}$ if and only if they differ by a constant spin assignment, so $B^1(P;\Z_q)$ is $13$-dimensional. As surmised, the quotient $H^1(P;\Z_q)$ is $15-13=2$-dimensional, so $\abs{H^1(P;\Z_q)}=q^2$.

			\subsection{Hyperlattice gauge theory} 
				\label{appendix:extended-probability}
				\noindent We can extend the random-cluster and Potts measures to ones on subcomplexes of arbitrary dimension. The \emph{$(i-1)$-dimensional Potts (hyper)lattice gauge theory $\nu$} on a finite cubical complex $X$ with parameters $p = 1-e^{-\beta}$ and integer $q$ is the Gibbs distribution on $C^{i-1}(X; \Z_q)$ induced by the Hamiltonian \[ \H(f) = -\sum_{x \in X} \1_{(\delta^{i-1}f)(x)=0}, \] where $f \in C^{i-1}(X; \Z_q)$, and $x$ are $i$-cells of $X$. Thus, \[ \nu(f) = \frac{1}{\PLGTpartition} e^{-\beta \H(f)}, \] where $\PLGTpartition$ is the normalizing constant \[ \PLGTpartition = \sum_{f^*} e^{-\beta \H(f^*)}, \] summing over all $(i-1)$-cochains $f^*$.
				
				For a finite cubical complex $X$, the \emph{$i$-dimensional plaquette random-cluster measure $\mu$} with parameters $p \in [0,1]$ and integer $q$ is the probability distribution \[ \mu(P) = \frac{1}{\PRCMpartition} \ p^{|P|}(1-p)^{|X|-|P|} \abs{H^{i-1}(P;\Z_q)} \] over $i$-subcomplexes $P \subseteq X$ that contain all cells of dimension $(i-1)$ or lower. The normalizing constant $\PRCMpartition$ is \[ \PRCMpartition = \sum_{P^*} p^{|P^*|}(1-p)^{|X|-|P^*|} \abs{H^{i-1}(P^*;\Z_q)} \] summing over all $i$-subcomplexes $P^* \subseteq X$ containing all cells of dimension $(i-1)$ or lower. When $q$ is prime, we can re-define the distribution and normalizing constant to \[ \mu(P) = \frac{1}{\PRCMpartition} \ p^{|P|}(1-p)^{|X|-|P|}q^{\betti_{i-1}(P)} \] and \[ \PRCMpartition = \sum_{P^*} p^{|P^*|}(1-p)^{|X|-|P^*|}q^{\betti_{i-1}(P^*)},\] where \[ \betti_{i-1}(Q) = \rank(H_{i-1}(Q, \Z_q))\] is the \emph{$i$\th Betti number} of $Q \subseteq X.$ \cite{duncan2023sharp,bernoulli-cell-complexes}

                The plaquette Swendsen--Wang and plaquette invaded cluster algorithms immediately generalize to these systems. The $i$-dimensional PRCM also satisfies a duality property: if $P$ is distributed as the $i$-dimensional PRCM with parameters $p$ and $q$, then the dual complex $P^{\bullet}$ obtained by including cells in $\left(\Z^d\right)^\bullet$ that do \emph{not} intersect a cell of $P$ is a $(d-i)$-dimensional PRCM with parameters \[ p^{\bullet}\paren{p,q} = \frac{p}{p+q\paren{1-p}} \] and $q$ after boundary conditions are correctly accounted for. See~\cite{prcm-24} for a precise statement. 

                Continuing the discussion of Section~\ref{section:duality}, suppose that $P$ is distributed as the $i$-dimensional PRCM. The distribution of $P$ is unchanged by the following operation:
                
                \begin{headlessboxedalgorithm}
                	\begin{enumalgorithm}
	                    \item Let $P^{\bullet}$ be the dual complex of $P.$
	                    \item Sample a uniform random cocycle $f$ from $Z^{d-i-1}\paren{P^{\bullet};\Z_q}$.
	                    \item Let $P^{\bullet}_{\mathrm{new}}$ include the dual $(d-i)$-faces on which $\delta^{i-1} f$ vanishes independently with probability $p^{\bullet}\paren{p,q}$. 
	                    \item Let $P_{\mathrm{new}}$ be the dual complex of $P^{\bullet}_{\mathrm{new}}$.
	                \end{enumalgorithm}
                \end{headlessboxedalgorithm}

                To restate this operation without duality, we use the \emph{discrete Hodge star operator} \[ \ast:C^{d-i-1}\paren{\paren{\Z^d}^{\bullet};\Z_q}\to C_{i+1}\paren{\Z^d;\Z_q}.\] Towards that end, let $\sigma^\bullet \in (\Z^d)^\bullet$ be the ${(d-i-1)}$-cell dual to the $(i+1)$-cell $\sigma \in \Z^d$, and define $f_{\sigma^\bullet} \in C^{d-i-1}((\Z^d)^\bullet;\Z_q)$ by setting $f_{\sigma^{\bullet}}\paren{\sigma}=\1_{\sigma=\sigma^{\bullet}}$ and extending linearly. The $\ast$ operator is the linear extension of the map $f_{\sigma^{\bullet}}\mapsto \sigma$, where $\partial \ast =\ast \delta$.
                
                Now, if $f$ is a uniformly random element of $Z^{d-i-1}\paren{P^{\bullet};\Z_q}$, the previous identity implies that $\partial \ast f$ is an $i$-cycle supported on $P$. Assuming the underlying complex is simply connected, this cycle is a uniform random element of $Z_i\paren{P;\Z_q}$. Putting everything together, the $i$-cells of $P_{\mathrm{new}}$ are found by starting with the \emph{trace} of this uniform random $i$-cycle --- the set of $i$-cells appearing in it with a non-zero coefficient --- and adding additional $i$-cells independently with probability $p^{\bullet}\paren{p,q}$. In all, we have sampled the PRCM via the generalized Loop--Cluster coupling defined in Section 4 of \cite{hansen2025general}.

	\section{Dynamic critical exponents}\label{appendix:exponents}
		\begin{table*}[b]
	\centering
	\setlength{\arrayrulewidth}{1pt}
		\begin{tabular}{c|c>{\columncolor[gray]{0.8}}c|c>{\columncolor[gray]{0.8}}c}
			$N$ & $\tau_{\mathcal N}$ & $\tau_{\mathcal N}$ \cite{ossola-sokal} & $\tau_{\mathcal E}$ & $\tau_{\mathcal E}$ \cite{ossola-sokal} \\ \hline
			$4$ & $2.0096 \pm 0.2742$ & $2.1169 \pm 0.0018$ & $2.3419 \pm 0.3041$ & $2.3697 \pm 0.0021$ \\
			$6$ & $2.4980 \pm 0.3181$ & $2.7257 \pm 0.0026$ & $2.8717 \pm 0.3529$ & $3.0618 \pm 0.0031$ \\
			$8$ & $3.0887 \pm 0.3735$ & $3.2298 \pm 0.0033$ & $3.5027 \pm 0.4125$ & $3.6496 \pm 0.0040$ \\
			$12$ & $3.7788 \pm 0.4355$ & $4.0638 \pm 0.0047$ & $4.2931 \pm 0.4783$ & $4.6314 \pm 0.0058$ \\
			$16$ & $4.2462 \pm 0.4757$ & $4.7701 \pm 0.0027$ & $4.8360 \pm 0.5214$ & $5.4588 \pm 0.0033$ \\
			$24$ & $5.3386 \pm 0.5612$ & $5.9567 \pm 0.0083$ & $6.3495 \pm 0.6384$ & $6.8408 \pm 0.0102$ \\
			$32$ & $6.5786 \pm 0.6560$ & $6.9303 \pm 0.0074$ & $7.4935 \pm 0.7220$ & $7.9625 \pm 0.0090$ \\
			$48$ & $7.4909 \pm 0.7219$ & $8.5612 \pm 0.0073$ & $8.7650 \pm 0.8142$ & $9.8308 \pm 0.0090$ \\
		\end{tabular}
	\caption{Integrated autocorrelation times ($\pm$ one standard deviation). Shaded columns are high-precision values from \cite{ossola-sokal}.}
	\label{table:autocorrelation-3d}
\end{table*}
		\noindent To lay out the statistical strategy for estimating dynamical critical exponents, we briefly summarize Appendix C of~\cite{madrasPivot1988} and Section 3 of~\cite{ossola-sokal}.
        
        Our quantities of interest are the \emph{plaquette occupation} (or \emph{plaquette density}) \[ \mathcal N(P,f) = |P| \] and (minus) the \emph{total energy} \[ \mathcal E(P,f) = \sum_{x \in X} \1_{\delta^1 f(x) = 1} \] for compatible pairs $(P,f)$ produced by the plaquette Swendsen--Wang (PSW) algorithm on systems of various scale. For each observable $Z$, we define natural estimators for the \emph{unnormalized autocorrelation time} \[ \rho_Z(t) = \frac{1}{n - |t|} \sum_{i=1}^{n-|t|} (Z_i - \overline Z)(Z_{i+|t|} - \overline Z), \] where $n$ and $\overline Z$ are the sample size and mean of $Z$, respectively; the \emph{normalized autocorrelation time} \[ \overline \rho_Z(t) = \rho_Z(t)/\rho_Z(0) = \rho_z(t)/\Var(Z) \] for all $0 \leq t \leq n$; and the \emph{integrated autocorrelation time} \[ \tau_Z(n) = \frac 12 \sum_{t=-n}^n \overline \rho_Z(t). \] All quantities are taken in equilibrium~\cite{ossola-sokal,madrasPivot1988}.
        
        As the $\overline \rho_z(t)$ are biased estimators, their error accumulates in $\tau_Z(n)$ as $t \to n$, making $\tau_Z(n)$ a poor estimate of the true integrated autocorrelation time. To mitigate this effect, we use the ``auto-windowing'' algorithm from~\cite{ossola-sokal,madrasPivot1988} that chooses a cutoff $M \ll n$ for the upper limit of the sum in the definition of $\tau_Z(n)$: given a constant $c \approx 6$, find the largest integer $M$ such that $c \tau_Z(M) \leq M$. Intuitively, this procedure detects the time-step at which $\tau_Z$ stops increasing according to the ``signal'' of the expected covariances $\rho(t)$, and starts varying according to their ``noise.'' We then report the estimated integrated autocorrelation time as $\tau_Z \coloneqq \tau_Z(M)$.

        Introducing this cutoff also gives us a clean expression for approximating the variance of the estimator $\tau_Z$: specifically, \[ \Var(\tau_Z) \approx \frac{2(2M + 1)}{n} \mathfrak T_Z^2, \] where $\mathfrak T_Z$ is the true integrated autocorrelation time. Here, we further assume that the random process is ``almost-Gaussian,'' in the sense that the covariances $\Cov(\overline \rho(i), \overline \rho(j))$ (and thus $\Var(\tau_Z)$) have no terms reliant on statistics of order greater than $2$. In practice, we compute the variance of $\tau_Z$ by replacing $\mathfrak T_Z$ with $\tau_Z(M)$ in the shortcut expression above. We then report the estimated integrated autocorrelation times as $\tau_Z(M) \pm \sqrt{\Var(\tau_Z)}$, giving $\tau_Z$ one-standard deviation error bars --- that is, the expression for $\tau_Z$ is a $68\%$ confidence interval~\cite{ossola-sokal,madrasPivot1988}.

        To fit the $\tau_Z$ at scale $N$ to a power law ${\tau_z \sim A \cdot N^z}$, we use the \curvefit function from the SciPy Python library. It uses a trust region reflective solver to obtain a fit, and (alongside parameter estimates for $A$ and $z$) outputs a $2 \times 2$ matrix $C$ of covariance estimates~\cite{virtanenSciPy2020}. We report the estimated dynamical critical exponents as $z \pm \sqrt{\Cov(z,z)} = z \pm \sqrt{C[z,z]}$, again specifying a one-standard deviation error bar (a $68\%$ confidence interval).
		
		For quality assurance purposes, we use \ATEAMS to partially replicate Ossola and Sokal's work in \cite{ossola-sokal} by computing the above estimators at the critical point of the $1$-dimensional Ising model on $\T^3_N$. Table \ref{table:autocorrelation-3d} compares our estimates for $\tau_{\Energy}$ and $\tau_{\Occupancy}$ to the high-precision ones found by Ossola and Sokal in \cite{ossola-sokal}. We found that our calculations consistently under-estimate the high-precision ones, which we largely attribute to fewer and (significantly) shorter runs. The estimates in \cite{ossola-sokal} report $z_{\Energy} = z_{\Occupancy} = 0.459 \pm 0.005$ on lattice sizes between $N=96$ and $N=256$; our estimates report $z_{\Energy} = z_{\Occupancy} = 0.499 \pm 0.02$ on the lattice sizes listed in Table \ref{table:autocorrelation-3d}.
		

	\section{\ATEAMS}\label{appendix:ATEAMS}
		\noindent \ATEAMS is an open-source library written in C++, Cython, and Python. It implements Algorithm \ref{algorithm:psw}, Algorithm \ref{algorithm:pic}, and Glauber dynamics in the \code{SwendsenWang}, \code{InvadedCluster}, and \code{Glauber} classes, respectively. \ATEAMS also provides narrow Python bindings to \SparseRREF and \SpaSM. Matrices are stored in COO (``coordinate-list'') format, and so require space linear in the number of cells; these matrices are tied to a \code{Cubical} complex object that can be written to and read from a compressed file format for re-use by any of the models above. To give a sense of these objects' sizes, the boundary matrix $\partial_2$ for $\T^4_{40}$ has $15{,}360{,}000$ columns.
		
		\subsection{Computing persistence}\label{appendix:ATEAMS:optimization}

			\noindent When computing persistence over $\Z_q$ for $q=2$, we use the \PHAT library \cite{phat}. For $q>2$, we change tack: by duality (of linear maps), we know that the map \[j: H_i(P_t, \Z_q) \to H_i(X, \Z_q)\] induced by inclusion is nontrivial whenever its dual \[j^*: H^i(X; \Z_q) \to H^i(P_t; \Z_q)\] is.\footnote{Given a matrix $A$, this is equivalent to $\rank(A) = \rank(A^\top)$.} If $P_t \subseteq X$ is a percolation subcomplex and $v$ a vector with entries corresponding to $i$-plaquettes of $X$, let $v_t$ be the vector containing all entries of $v$ corresponding to $i$-plaquettes in $P_t$; similarly, let $\delta^k_t$ be the $k$\th coboundary matrix with rows (and columns, if applicable) corresponding to $i$-plaquettes in $P_t$.
				
			\vspace{0.75em}
			\begin{proposition}\label{proposition:duality}
				There exists a $\gamma \in H_i(P_t, \Z_q)$ such that $j(\gamma) = e$ iff $e^*_t$ is \emph{not} in the image of $\delta^{i-1}_t$. 	
			\end{proposition}
			\vspace{-0.5em}
			\begin{proof}
				Note that under row deletion (from $e^*$) and column deletion (from $\delta^i$), $e^*_t$ is always in $\ker(\delta^i_t)$. To prove the forward direction, suppose $e^*_t$ is in $\im(\delta^{i-1}_t)$, so $j^*(e^*) = 0$. Thus $j^*(e^*) = e^* \circ j$ is the zero map and, as $e^*$ is nonzero, there cannot exist a $\gamma \in H_d(P_t)$ such that $j(\gamma) = e$. For the backward direction, suppose $e^*_t$ isn't in $\im(\delta^{i-1}_t)$, so $j^*(e^*) = e^* \circ j$ is nonzero. Since $e^*$ takes $e$ to $1$ and everything else to $0$, there must be a $\gamma \in H_d(P_t)$ for which $(e^* \circ j)(\gamma) = 1$, implying $j(\gamma) = e$.
			\end{proof}
			
			\noindent Because the ambient cubical complex $X$ is fixed, we can pre-compute a basis $\mathcal B$ for $H_i(X, \Z_q)$ --- and thus a dual basis $\mathcal B^*$ for $H^i(X; \Z_q)$ --- and use the backward implication of Proposition \ref{proposition:duality} to determine when $\rank(j)$ attains a desired value $\mathcal C$. Proceeding as a binary search by setting the initial search window as $\left[0, |X^i|\right]$, $t \coloneqq \nicefrac{|X^i|}{2}$, and $A_t \coloneqq \left[ \mathcal B^*_t \mid \delta^{i-1}_t \right]$,

\begin{headlessboxedalgorithm}
	\begin{enumalgorithm}
		\item Set $r_t \coloneqq -1$. While $r_t \neq \mathcal C$,
		\begin{enumalgorithm}
			\item Compute $r_t \coloneqq \rank(A_t)-\rank(\delta^{i-1}_t)$, and $r_{t+1}$ similarly.
			\item If $r_t < \mathcal C = r_{t+1}$, go to Step \ref{algorithm:binary:report}.
			\item If $r_{t+1} < \mathcal C$, search the upper half of the current window; if $r_t = \mathcal C \leq r_{t+1}$, search the lower half of the current window.
		\end{enumalgorithm}
		\item\label{algorithm:binary:report} Report $t+1$.
	\end{enumalgorithm}
\end{headlessboxedalgorithm}

		\subsection{Performance}\label{appendix:ATEAMS:performance}
			\vspace{0.25em} \noindent Below are performance statistics for various configurations of the \code{SwendsenWang} and \code{InvadedCluster} models. We perform $100$ iterations of each configuration on \Pangolin, a Dell Precision 5280 workstation with an 18-core Intel Xeon W-2295 processor clocked to 1.3GHz. We will continue to update the tables with data from larger simulations. The columns are:
			
			\begin{descriptor}
				\item[$N$] Scale (side length) of the torus $\T^d_N$.
				\item[Plaq] Total number of plaquettes in $\T^d_N$.
				\item[Field] Finite field over which computations were performed.
				\item[s/it] Seconds per iteration.
			\end{descriptor}
			
			\noindent \code{SwendsenWang}'s speed increases with $q$, as \SparseRREF tends to handle fields of larger order more quickly. In practicality, experiments with \code{SwendsenWang} are efficient and easily performed on a laptop computer.
						
			\clearpage
			\onecolumn
			\subsection{Swendsen--Wang}
				\begin{figure*}[h!]
	\setstretch{1.25}
	\centering
	\small
	\begin{tikzpicture}
		\tikzset{
			header/.style={anchor=south,font={\small\itshape},inner sep=0.15in},
			img/.style={inner sep=0pt}
		}
		\node[img, anchor=east] (sw-2) at (0,0) {\input{tables/SwendsenWang.ReducedKernelSample.2.tex}};
		\node[header] at (sw-2.north) {Performance in $\T^2_N$.};
		
		\node[img, anchor=north west] (sw-4) at ($(sw-2.north east)+(0.5in,0)$) {\input{tables/SwendsenWang.ReducedKernelSample.4.tex}};
		\node[header] at (sw-4.north) {Performance in $\T^4_N$.};
	\end{tikzpicture}
\end{figure*}

			\subsection{Invaded-cluster}
				\begin{figure*}[h!]
	\setstretch{1.25}
	\centering
	\begin{tikzpicture}
		\tikzset{
			header/.style={anchor=south,font={\small\itshape},inner sep=0.15in},
			img/.style={inner sep=0pt}
		}
		\node[img, anchor=east] (sw-2) at (0,0) {\input{tables/InvadedCluster.Persistence.2.tex}};
		\node[header] at (sw-2.north) {Performance in $\T^2_N$.};
		
		\node[img, anchor=north west] (sw-4) at ($(sw-2.north east)+(0.4in,0)$) {\input{tables/InvadedCluster.Persistence.4.tex}};
		\node[header] at (sw-4.north) {Performance in $\T^4_N$.};
	\end{tikzpicture}
\end{figure*}

\end{document}